\documentclass[a4paper,11pt]{article}
\pdfoutput=1
\usepackage{jcappub}
\usepackage[utf8]{inputenc}
\usepackage{booktabs}
\usepackage{rotating}
\usepackage{array}
\usepackage{tabularx} 
\usepackage{graphicx}
\usepackage{dcolumn}
\usepackage{bm}
\usepackage{hyperref}
\usepackage{amsmath,amsfonts,amssymb,slashed,upgreek,color,amscd,cancel,footnote}
\usepackage{multirow}
\usepackage{subfigure}

\newcommand{\be}{\begin{equation}}
\newcommand{\ee}{\end{equation}}
\newcommand{\beq}{\begin{equation}}
\newcommand{\eeq}{\end{equation}}
\newcommand{\bea}{\begin{eqnarray}}
\newcommand{\eea}{\end{eqnarray}}

\newcommand{\R}{{}^{(3)}\!R}

\newcommand{\pu}{p^{(1)}}

\newcommand{\gs}{\gamma_{\rm MG}}

\newcommand{\Mb}{M}
\newcommand{\Lb}{\lambda}
\newcommand{\cb}{{\cal C}}

\urldef\mgcamb\url{http://www.sfu.ca/~aha25/MGCAMB.html}
\urldef\planckProducts\url{http://pla.esac.esa.int/pla/aio/planckProducts.html}

\title{Constraints on modified gravity from Planck 2015: \\ when the health of your theory makes the difference.}

\author[a,1]{Valentina Salvatelli,\note{Corresponding author.}}
\author[a]{Federico Piazza,}
\author[a]{Christian Marinoni.}
\affiliation[a]{Aix Marseille Universit\'e, CNRS, CPT, UMR 7332, 13288 Marseille,  France.}
\emailAdd{Valentina.Salvatelli@cpt.univ-mrs.fr}
\emailAdd{Federico.Piazza@cpt.univ-mrs.fr}
\emailAdd{Christian.Marinoni@cpt.univ-mrs.fr}

\abstract{ We use the effective field theory of dark energy (EFT of DE) formalism to constrain dark energy models belonging to the \emph{Horndeski} class with the recent Planck 2015 CMB data. The space of theories is spanned by a certain number of parameters determining the linear cosmological perturbations, while the expansion history is set to that of a standard $\Lambda$CDM model.  We always demand that the theories be free of fatal instabilities. Additionally, we consider two optional conditions, namely  that  scalar and  tensor perturbations propagate with subliminal speed. Such criteria severely restrict the allowed parameter space and are thus very effective in shaping the posteriors. As a result, we confirm that no theory performs better than $\Lambda$CDM when CMB data alone are analysed. Indeed, the healthy dark energy models  considered here are not able to reproduce those phenomenological behaviours of the effective Newton constant and gravitational slip parameters that, according to previous studies, best fit the data.}

\begin{document}
\maketitle
\flushbottom

\section{Introduction} \label{sec:intro}
Understanding the  origin of the present acceleration of the universe  is a key challenge for cosmology.
Recent progress in the analysis of the Cosmic Microwave Background \cite{Ade:2015cp} has significantly 
strengthen the case for the so called $\Lambda$CDM model, in which the Einstein field equations are supplemented by a cosmological constant, and the dominant matter specie  is cold dark matter. 
Besides fixing with high precision the parameters of the standard model, in some cases to sub-percentage level, CMB  data  impose stringent 
constraints also on new non-standard physics.  A large class of  dark energy scenarios, in which cosmic acceleration results from a time varying dark energy fluid or modifications in  the action of the  gravitational field,  are now shown to be in conflict with observational evidences \cite{Betoule:2014frx, Bel:2014awa}.

Beside its theoretical simplicity, a most compelling  virtue of $\Lambda$CDM is its ability to reproduce the observed cosmic expansion history.  However,  well beyond the behaviour of the homogeneous Universe as a whole, there are specific aspects of the evolution of the structures it contains, such as the way density fluctuations grow and deflect photons via the lensing mechanism, which still escape full  understanding.  
Indeed the six-parameters $\Lambda$CDM ``calibrated'' by  Planck at high redshift
seems to predict that structures grow faster in time and are more abundant in space than  actually measured by galaxy surveys at $z\lesssim1$. 
This is illustrated by the fact that the $rms$ density fluctuations on the scale of $8 h^{-1}$ Mpc---extrapolated from CMB data under the assumption of a $\Lambda$CDM universe governed by general relativity---is larger than the value effectively measured by means of a variety of galaxy observables,  such as cluster counts \cite{Vikhlinin:2008ym, Ade:2015fva, Ilic:2015rna}, lensing \cite{Heymans:2013fya, Battye:2013xqa, Ade:2015cp, Raveri:2015maa} and redshift space distortions \cite{Macaulay2013, delaTorre:2013rpa, Beutler:2013yhm, Samushia:2013yga, Anderson:2013zyy, Steigerwald:2014ava}. 

In this perspective, it is certainly interesting the indication of~\cite{Ade:2015mg}, confirmed by the following analysis of~\cite{DiValentino:2015bja}, of a  tension  between the $\Lambda$CDM model scenario for structure formation and the available data that could be explained in terms of modified gravity. As far as linear cosmological perturbations are concerned, it is possible to boil down the effects of most modified gravity models to two dimensionless functions~\cite{Zhao:2008bn,Pogosian:2010tj}:  the ratio between the gravitational coupling (as it appears in the Poisson equation) and the Newton constant, $\mu_{\rm MG}=G_{\rm eff}/G_N$, and the ratio between the two gravitational potentials $\gamma_{\rm MG}=\Psi/\Phi$. Since we can neglect anisotropic stress at late times, both quantities reduce to unity in the standard model.  Additionally the quantity $\Sigma= \mu_{\rm MG}(1+\gamma_{\rm MG})/2$, directly corresponding to the lensing potential, can be used to probe modified gravity. Anomalous values of these quantities are effectively reported in \cite{Ade:2015mg}, who found  $3\sigma$ evidence against the $\Lambda$CDM model when low-redshift probes are combined with CMB. 

It is certainly premature to interpret  these results as indication that the standard model of cosmology is  missing  some fundamental degree of freedom. Indeed, a strong statistical discrepancy arises only when galaxy weak lensing or redshift space distortions data are included---but the latter probes still lack the understanding of systematics of CMB experiments (\emph{e.g.} \cite{Bianchi:2012za, Kitching:2016hvn, Joudaki:2016mvz}). Nonetheless, there is much hope that statistical and systematic errors will be minimized and brought under control in the next generation of redshift galaxy surveys such as Euclid~\cite{Euclid}, DESI~\cite{DESI} or eBOSS~\cite{eBOSS}. 
While waiting for future observational  confirmation or disproval, it is worth investigating which, among the many  theoretical models, 
is best suited  for making sense of the observed discrepancies. 

Instead of using phenomenological parameterizations, we  propose here to describe deviations from the standard scenario directly in terms of ``constitutive parameters" of alternative gravitational theories. This is made possible  by a formalism that allows to describe disparate theoretical models of DE in a unified language. The \emph{effective field theory of dark energy} (EFT of DE), at least in its minimal version, allows to explore all dark energy and modified gravity models that contain one additional scalar degree of freedom~\cite{Creminelli:2008wc,EFTOr,Bloomfield:2012ff,GLPV,Bloomfield:2013efa,PV,Tsujikawa:2014mba,Gleyzes:2015rua} (see~\cite{eftcamb1,eftcamb2,eftcamb3} for a numerical implementation of this formalism). Adding another scalar~\cite{Gergely:2014rna} or a non-minimal coupling dark energy-dark matter~\cite{Gleyzes:2015pma} is also relatively natural in this framework. 

In this work we use EFT of DE to explore which modified gravity models are compatible with CMB temperature, polarisation and lensing power spectra. For definiteness, we will limit our analysis to those models that give perturbation equations containing up to two derivatives (\emph{Horndeski} models~\cite{horndeski}, that can be seen as generalizations~\cite{Deffayet:2009mn,Deffayet:2011gz} of \emph{galileon} models~\cite{NRT}). Our goal is twofold. On the one hand,  we want to  single out  specific MG models, in the Horndeski class, that are compatible with  data and ultimately assess, via a Bayesian analysis of their evidence,  whether these models are more likely than  the standard picture. 
By doing this we  aim at reproducing and  extending   preliminary analyses and results already presented in \cite{eftcamb2,Ade:2015mg,Bellini:2015xja}. 
On the other hand, the novelty of the paper is that we disentangle in our analyses the constraining power of data from that of the theory, \emph{i.e.} we  highlight which portion of the parameter space spanned by non-standard theory is excluded not because of tension with observations, but because no healthy physical model is allowed there.  We clearly show that the theory constraining power greatly helps in reducing the volume of the multidimensional parameter space that is statistically explored, as~\cite{pheno,pheno2} suggested. 

The paper is organised as follows. In Section~\ref{sec:theory} we recall the main elements of the EFT formalism and we describe the parametrization we adopt. In Section~\ref{sec:method} the method of analysis and the datasets are explained. In Section~\ref{sec:resultsEFT} we present the results in the space of parameters. In Section~\ref{sec:resultsOBS} we show some results directly in the space of observables. In Section~\ref{sec:concl} we draw our conclusions.

\section{EFT formalism and parametrization} \label{sec:theory}

The effective field theory of dark energy allows to describe a vaste range of dark energy models by using a limited number of time dependent couplings~\cite{Creminelli:2008wc,EFTOr,Bloomfield:2012ff,GLPV,Bloomfield:2013efa,PV}. In particular, here we focus on the large class of theories containing up to one scalar degree of freedom in addition to the metric field, and up to two derivatives in the equations of motion--- commonly defined as Horndeski theories. 
Upon use of the Friedmann equations, the relevant couplings can be reduced to a minimal set of truly independent functions and the split between background expansion history and perturbation quantities becomes complete~\cite{pheno,BS,Gleyzes:2014qga,Gleyzes:2014rba,pheno2}. 
While there is now a consensus on the power and the advantages of this formalism, there is no universal agreement on the conventions for the coupling functions yet. Here we use those of~\cite{pheno,pheno2}, that maintain a more direct link with the underlying theories, with respect to those of Ref.~\cite{BS,Gleyzes:2014rba}. For a dictionary between the two notations we refer the reader to App.~B of Ref.~\cite{Gleyzes:2014qga}. 

\subsection{Background expansion history}
One of the main advantages of the EFT formalism is the possibility of treating cosmological perturbations independently of the expansion history. 
As far as the latter is concerned, we fix the geometry of the Universe to that of a spatially flat  $\Lambda$CDM model. This is fully consistent with the present observational status of the equation of state parameter~\cite{Betoule:2014frx,Ade:2015cp,Aubourg:2014yra}.
The Hubble rate $H(z)$ as a function of the redshift is thus given, at late times, by 
\begin{equation}
H^2(z) = H_0^2\left[x_0 (1+z)^3 + 1-x_0\right]\, .
\end{equation}
 The only free parameter here is $x_0$. In a real $\Lambda$CDM model this quantity corresponds to the fractional matter density today. Here, $x_0$ is only a proxy for the geometry of the universe, which fixes its background expansion history. Indeed, by exploiting the \emph{dark degeneracy} discussed \emph{e.g.} in~\cite{Kunz:2007rk,pheno,pheno2,darkdeg}, one could consider an interesting mismatch between the actual, physical amount of non-relativistic matter as accounted for in the energy momentum tensor, $\Omega_m^0 = \rho_m(t_0)/(3 M^2_{\rm Pl} H_0^2$) and $x_0$. In~\cite{pheno,pheno2}, such mismatch was encoded in a parameter $\kappa$ different than unity. From now on, here we simply set 
 \begin{equation}
 \Omega_m^0 = x_0\, ,
 \end{equation}
and leave studies of the dark degeneracy for future work. 

\subsection{Non-minimal couplings: perturbation sector}
In order to completely specify the perturbation sector we need four functions of the time, corresponding to the four non-minimal couplings: $\mu(t), \mu_2(t), \mu_3(t)$ and $\epsilon_4(t)$. Along the  $\mu(t)$ direction in the coupling space we find Brans-Dicke (\emph{BD})-type theories, while $\mu_3$, appears in cubic galileon- and Horndeski-3 theories. They are both parameters with mass dimensions, typically of order Hubble. On the other hand, $\epsilon_4$ is a dimensionless order-one function of the time
present in galileon/Horndeski 4 and 5  Lagrangians.\footnote{This parameter is responsible for the anomalous gravitational wave speed $c_T \neq 1$ in theories of modified gravity, \emph{i.e.} $c_T^2 = 1/(1+\epsilon_4)$. In the paper~\cite{BPV}, by using binary pulsar data, its present value, $\epsilon_4(t_0)$,  has been constrained to more than $10^{-2}$ level.} 
From now on, we do not consider the function $\mu_2(t)$, which only affects the sound speed of the scalar fluctuations and that we thus set to zero.
In summary, the background and perturbation sectors are characterized in the approach we follow by one parameter and three functions of the time:
\begin{equation} \label{couplings}
\left\{\Omega_m^0, \ \mu(t),\ \mu_3(t),\ \epsilon_4(t) \right\}\, .
\end{equation}

In order to specify the time dependence of the couplings, 
it is convenient to promote the fractional matter density of the background to a time variable for the late Universe
\begin{equation}
\label{xdef}
x\ \equiv \ \frac{\Omega_m^0}{\Omega_m^0 + (1- \Omega_m^0) (1+ z)^{- 3 }}\, .
\end{equation}
In fact, $x$ detaching from 1 triggers the rising of the recent dark energy dominated phase. It seems thus convenient to parametrize the time behaviour of the coupling functions in~\eqref{couplings} with the following expansion:
\begin{align} \label{tay1}
\mu\left(x\right)\ &=\ (1-x) \left[p_1+\pu_1 \left(x -\Omega_m^0\right) \right ] H(x) \, ,\\[2mm]
\mu_3\left(x\right)\ & =\ (1-x)\, \left[p_3+\pu_3 \left(x -\Omega_m^0\right) \right ] H(x) \, ,\label{tay2}\\[2mm]
\epsilon_4\left(x\right)\ &=\ (1-x) \,\left[p_4+\pu_4 \left(x -\Omega_m^0\right) \right ] \, \label{tay3},
\end{align}
where the $p_i$ are order-one coefficients that we want to constrain with our analysis.
The above ansatz guarantees that the coupling functions go to zero at early times, and that all modified gravity effects are linked to the latest, dark energy dominated phase.

However, even with the non-minimal couplings switched off, and the background expansion history has been set identical to that of a $\Lambda$CDM model, dark energy could be \emph{physically} persistent at very early times, \emph{i.e.} present in the energy momentum tensor. Since we are fixing the expansion history, the only way for this to be the case is that its equation of state asymptotes to zero, thereby mimicking dark matter at the level of the background. In order to avoid this possibility,  we impose a constraint between the $p_i$ parameters,
\begin{equation} \label{constraint}
\pu_1 = \frac{p_1 \log (\Omega_m^0)-6\log\left(1 + (1- \Omega_m^0) p_4\right)}{1 -\Omega_m^0+ \Omega_m^0 \log (\Omega_m^0)}\, .
\end{equation}
We refer the reader to~\cite{pheno2} for a more throughout explanation of this constraint. 

\subsection{Viability conditions}
\label{sec:viabcond}
The theory that we are describing contains one scalar and two tensor degrees of freedom. The viability conditions that we demand at any time is that such degrees of freedom are not affected by ghosts or gradient instabilities. A gradient term appearing in the quadratic Lagrangian for the fluctuations with the wrong sign would imply exponential growth of Fourier modes of any comoving momentum $k$. The wrong sign in the \emph{time} kinetic term, on the other hand, would lead to ``\emph{ghost-like}" classical and quantum instabilities that are at least as serious~\cite{Cline:2003gs}. As we will show, these conditions alone significantly restrict the  parameter space that we are exploring. On top of these basic requirements that are always enforced, we consider in our analysis other two optional conditions, namely that the speed of propagation of scalar modes and tensor modes be not superluminal. Apart from the known causality problems related with the possibility of sending a signal faster than light, superluminal propagation has been argued to be incompatible with a consistent Lorentz invariant UV completion~\cite{Adams:2006sv}.
In summary, in this paper we will consider three main viability conditions:
\begin{align}\label{gradcon}
{\rm stable:}  \ &   \quad \text{absence of ghosts and gradient instabilities},\\[2mm] \label{sublum}
{\rm stable\ \ \&\ \ } c_s<1: \ & \quad \text{the above \emph{and} scalar propagation speed not superluminal },\\[2mm]\label{lastcon}
{\rm stable\ \ \&\ \ } c_s<1\ \ \&\ \ c_T<1: \ & \quad \text{the above \emph{and} tensor propagation speed not superluminal. } 
\end{align}

\subsection{MGCAMB with the EFT of DE parameters.}
\label{subsec.MGCAMB}
In our analysis, instead of solving the full set of linear perturbation equations for the couplings defined in~\eqref{tay1}-\eqref{tay3}, we encode the modifications of gravity in two functions of the time, $\mu_{MG}$ and $\gamma_{MG}$, following the approach implemented in the MGCAMB code~\cite{Zhao:2008bn,Hojjati:2011ix} and properly updating and modifying the public package\footnote{http://www.sfu.ca/~aha25/MGCAMB.html}. This method has the remarkable advantage of allowing a simpler numerical implementation while keeping a clear mapping between the $\mu_{MG}$-$\gamma_{MG}$ functions and the underlying EFT theory.

Although MGCAMB works in synchronous gauge, the form of the equations and the definition of the relevant quantities look more transparent in Newtonian-gauge, defined  
by the perturbed metric taking the form
\begin{equation} \label{newtonian} 
ds^2 = -(1+2\Phi)dt^2 + a^2(t) (1-2 \Psi) \delta_{i j} dx^i dx^j\, .
\end{equation}

The package MGCAMB evolves the standard conservation and Euler equations for the matter fields, implemented with other two equations, namely,
\begin{align} \label{pi}
-\frac{k^2}{a^2}\Phi\, =\, \mu_{\rm MG}(t,k) \, \frac{3 H^2}{2}  \left[ \Delta + 3 \left(1+ \frac{p}{\rho} \right) \sigma\right] \, , \\
\frac{k^2}{a^2}\left[\Psi - \gamma_{\rm MG}(t,k) \Phi\right] \, = \, \mu_{\rm MG}(t,k) \, \frac{9 H^2}{2}\left(1+ \frac{p}{\rho} \right)\sigma\, .
\end{align}
In the above, $\Delta = \delta - 3 H (\rho + p) v$ is the comoving density perturbation, $\sigma$ the anisotropic stress, negligible at late times, and $\mu_{\rm MG}$ and $\gamma_{\rm MG}$ are generally functions of both the time $t$ and the comoving scale $k$.
Note that in the differential equations integrated by MGCAMB the scalar degree of freedom is absent, so its effects must be encoded in these two functions. As summarized in Ref.~\cite{pheno2}, a  closed form for $\mu_{\rm MG}$ and $\gamma_{\rm MG}$ can be derived in a rather simple way, by retaining the spatial gradient terms in the Newtonian gauge action and neglecting both mass terms and time derivative terms. This is the essence of the \emph{quasi-static approximation}, valid at distances shorter than the sound horizon $c_s H^{-1}$.
In this approximation, $\mu_{\rm MG}$ and $\gamma_{\rm MG}$ only depend on $t$.
In the EFT of DE formalism, they have been derived and discussed, \emph{e.g.},  in~\cite{GLPV,pheno,BS,Gleyzes:2014rba,pheno2},\footnote{When using the results of~\cite{pheno2}, one should keep in mind that, in that notation, $\mu_{\rm MG} = \kappa G_{\rm eff}/G_N$.}.
In our analysis, we have fed MGCAMB with the EFT expressions of  $\mu_{\rm MG}$ and $\gamma_{\rm MG}$ quoted in the Appendix in eqs.~\eqref{geff} and~\eqref{postn}. One can verify that, with our parameterization~\eqref{tay1}-\eqref{tay3},~\eqref{constraint}, $\mu_{\rm MG}$ and $\gs$ go to one at early times. 

We expect that the quasi-static approximation behind our approach may introduce some discrepancies with respect to the integration of the full set of linear equations. While a full comparison of the two approaches is not trivial and it is left for future work, we extrapolated from the comparison between MGCAMB and EFTCAMB in the particular case of $f(R)$ theories \cite{eftcamb2} that an error of at most $10\%$ at $l=2$ arises for theories with $c_s^2 \sim 1$. Theories with lower $c_s^2$ might presents higher discrepancies at low-l and for high deviations from general relativity due to the worsening of the quasi-static approximation. However, as we will see in Secs.~\ref{sec:resultsEFT} and~\ref{sec:resultsOBS} below, our posteriors are mainly driven by the viability conditions, being therefore mildly affected by the effects of this approximation.

\begin{figure*}
\centering
\includegraphics[scale=0.37]{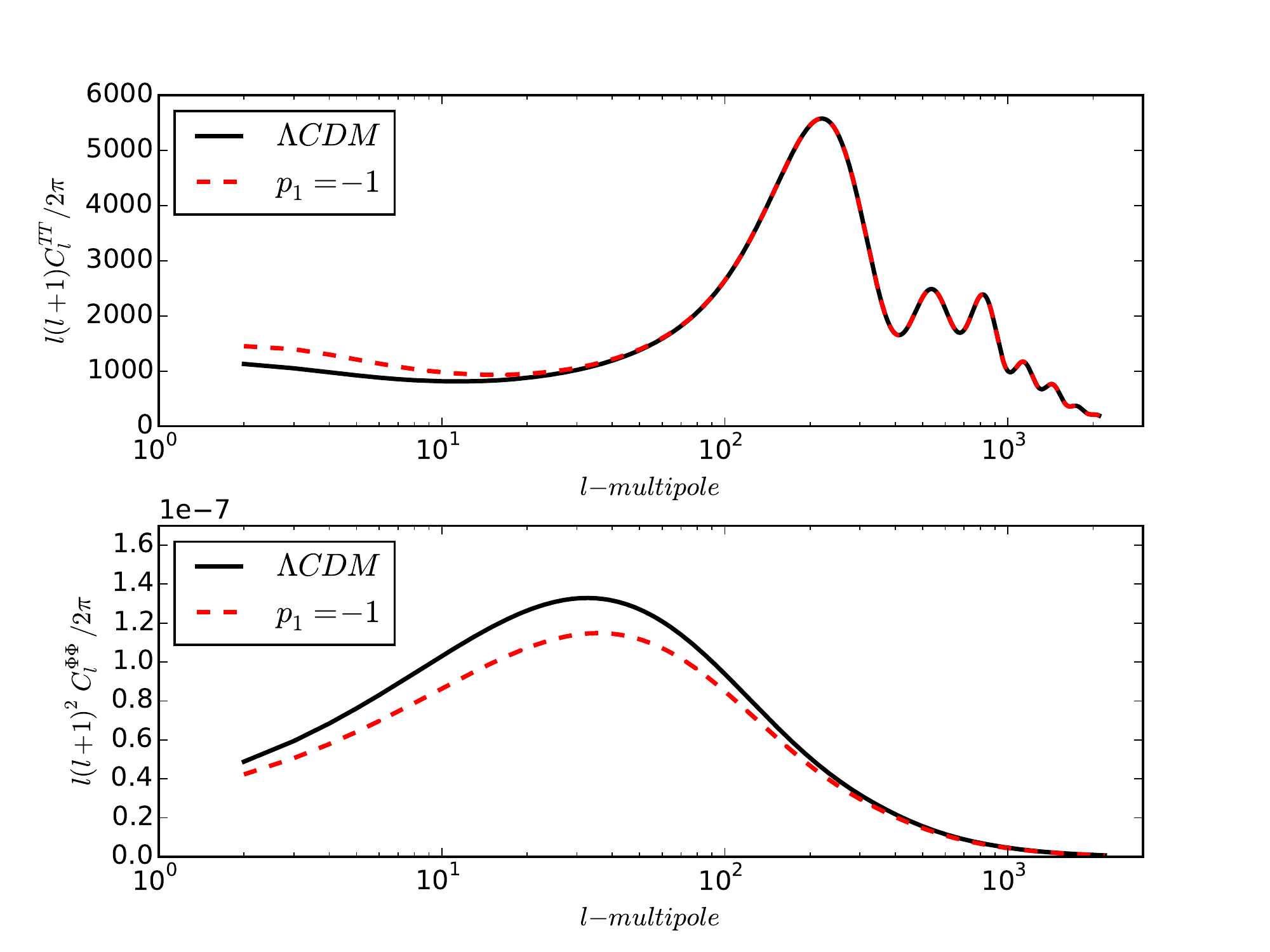}
\includegraphics[scale=0.37]{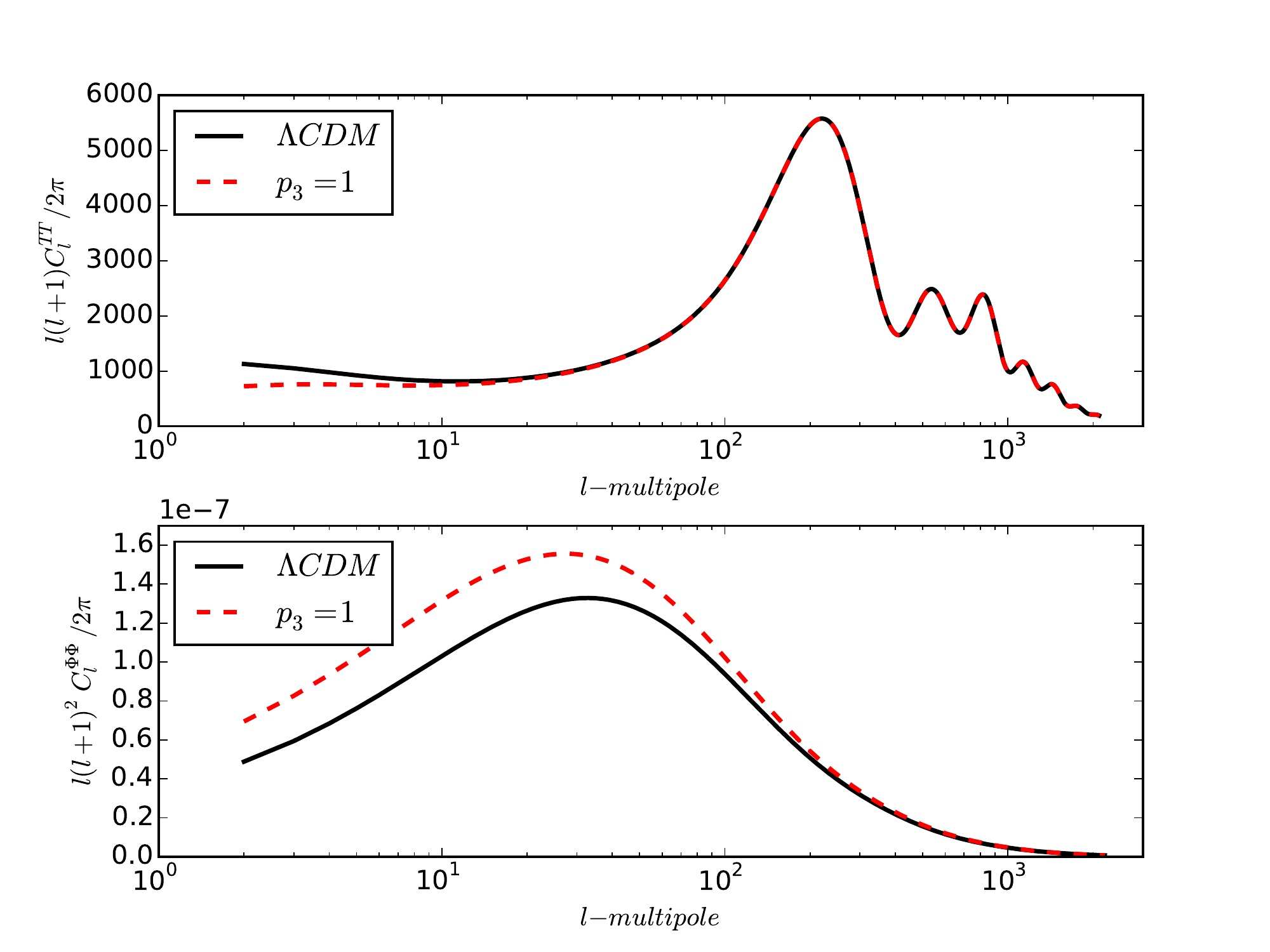}
\includegraphics[scale=0.37]{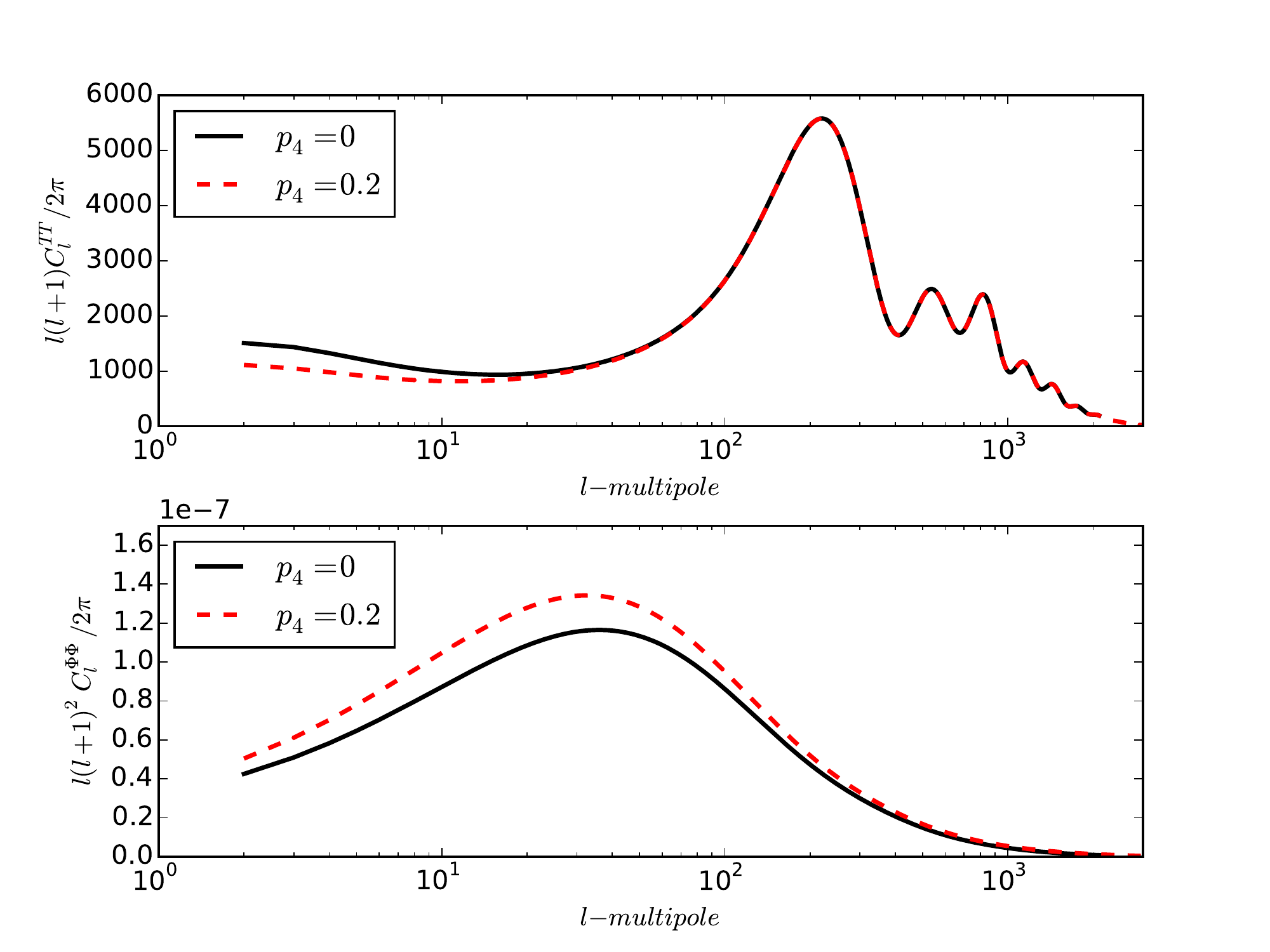}
\caption{\label{fig:spectra} Effects of the EFT couplings on the temperature and lensing CMB spectra. For the top figures we have switched on, in turn, the $p_1$ or $p_3$ parameter while keeping the others to zero. For the bottom figure, we have switched on the $p_4$ parameter, while keeping $\pu_3$, $\pu_4$ to zero and $p_1$ and $p_3$ fixed in a stable configuration. The signs have been chosen on the basis of stability. Note that a negative $p_1$ tends to be compensated by a positive $p_3$, which explains the degeneracy of Fig.~\ref{fig:2Dconstraints-ordzero} below, second panel. On the other hand, the effects of $p_3$ show up only at very low $l$, which explains why the likelihood is not very sensitive to this parameter (see Fig.~\ref{fig:posteriors-ordzero} and Panels 1 and 3 of Fig.~\ref{fig:2Dconstraints-ordzero}).}  
\end{figure*}

The effects of the different EFT parameters on temperature and lensing CMB power spectra are depicted in Fig.~\ref{fig:spectra}.

\section{Method of analysis and data} \label{sec:method}
The aim of the analysis is to simultaneously evaluate the constrains on the set of standard cosmological parameters $\Omega_{\rm b}h^2$, $\Omega_{\rm c}h^2$, $\theta$, $\tau$, $n_{\rm s}$, $A_{\rm s}$, that define a flat universe with a $\Lambda CDM$ background history, plus the $p_i$ coefficients that encode the dark energy/modified gravity effects, as described in \eqref{tay1}-\eqref{tay3}. 
In this respect, notice that  $\Omega_{\rm b}h^2$ is the baryon energy density, $\Omega_{\rm c}h^2$ is the cold dark matter energy density, $\theta$ is the ratio of the sound horizon to the angular diameter distance at the decoupling time, $\tau$ is the optical depth to reionization, $n_{\rm s}$ is the scalar spectral index and $A_{\rm s}$ is the amplitude of the primordial scalar perturbation spectrum, at $k = 0.05 Mpc^{-1}$. Deviations from standard cosmology in the neutrino sector are not considered in the following analysis. Therefore, the relativistic number of degrees of freedom parameter is fixed to $N_{eff} = 3.046$ and the total neutrino mass to $\sum m_{\nu}$ = 0.06eV. Notice that the $\Omega_m^0$ parameter in \eqref{couplings} corresponds to $(\Omega_{\rm b}h^2+\Omega_{\rm c}h^2)/h^2$.

In this analysis, we focus in this analysis on the most recent CMB data from the Planck experiment \cite{Adam:2015rua,Aghanim:2015xee}. In particular we include in our datasets the temperature high-$l$ power spectra from the 100, 143, 143x217 and 217 GHz channels (PlikTT likelihood) and the temperature and polarization spectra at low-l described in \cite{Aghanim:2015xee}, that includes Planck observations at low and high frequency channels, WMAP observations between 23 and 94 GHz \cite{bennett2012} and measurements at 408MHz from \cite{haslam1982}. We also include the information on CMB lensing from the trispectrum. We refer to this combination of datasets as PLANCK.\footnote{Notice that the high-$l$ polarization dataset is not included as it is insensitive to the EFT parameters and it does not improve the constraints.}

As already explained in Sec.~\ref{subsec.MGCAMB}, we explore the constraints on this set of parameters by computing the CMB observables with a MGCAMB code properly modified to include the EFT parametrization.

The public available CosmoMC package \cite{Lewis:2002ah,Lewis:2013hha}, version July2015, is used to explore the parameter space with the Monte Carlo Markov Chain method. The Gelman and Rubin method is used to set the convergence of the chains, requiring $R-1<0.03$.

We consider two different extensions of the standard model. The {\it 3D-}Model, that corresponds to a minimal extension with one free parameter for every non-minimal coupling. And the {\it 5D-}Model that, by adding a term in the Taylor expansion~\eqref{tay1}-\eqref{tay3} of the coupling functions, gives more freedom to the functional space. In summary, the two models are characterized by the following sets of EFT free parameters:
\begin{align}
{\rm {\it 3D-}Model}: \qquad \quad &\qquad \left\{p_1, \ p_3, \ p_4\ \right\} \\[1.6mm]
{\rm {\it 5D-}Model}: \qquad \quad &\qquad \left\{p_1, \ p_3, \ p_4, \ \pu_3\!\!, \ \pu_4 \right\}\, .
\end{align}
In the first model $\pu_3$ and $\pu_4$ are set to zero. 

\section{Results: Constraints on EFT parameters} \label{sec:resultsEFT}
\begin{figure*}
\centering
\includegraphics[scale=0.39]{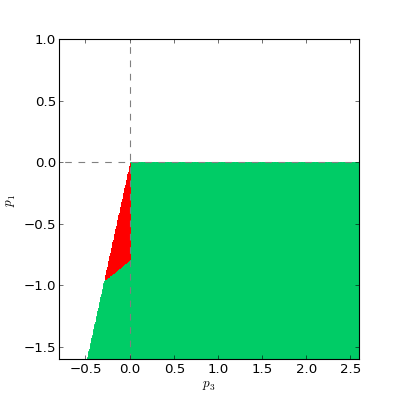}
\includegraphics[scale=0.39]{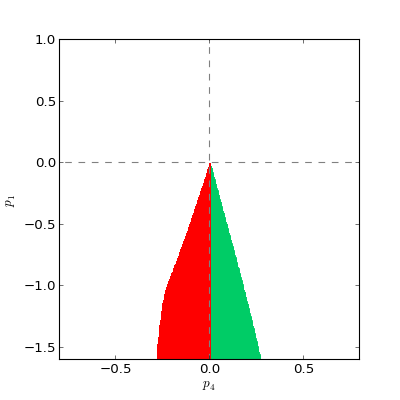}
\includegraphics[scale=0.39]{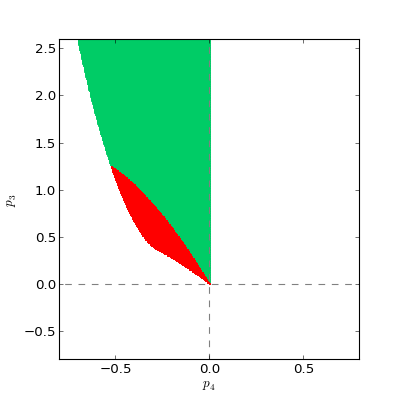}
\includegraphics[scale=0.58]{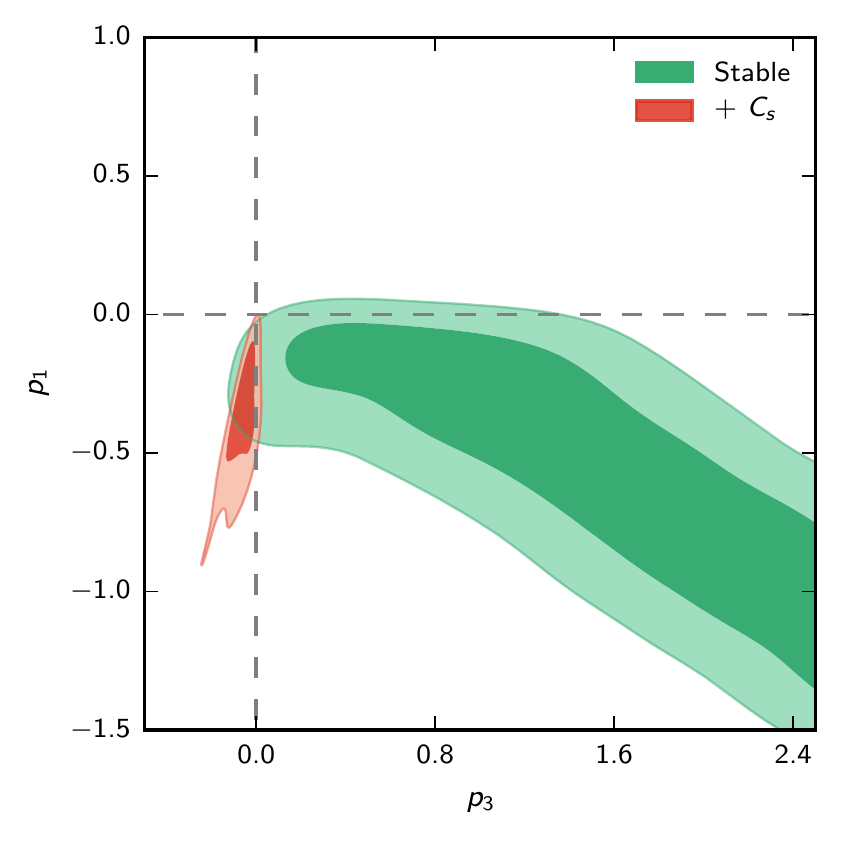}
\includegraphics[scale=0.58]{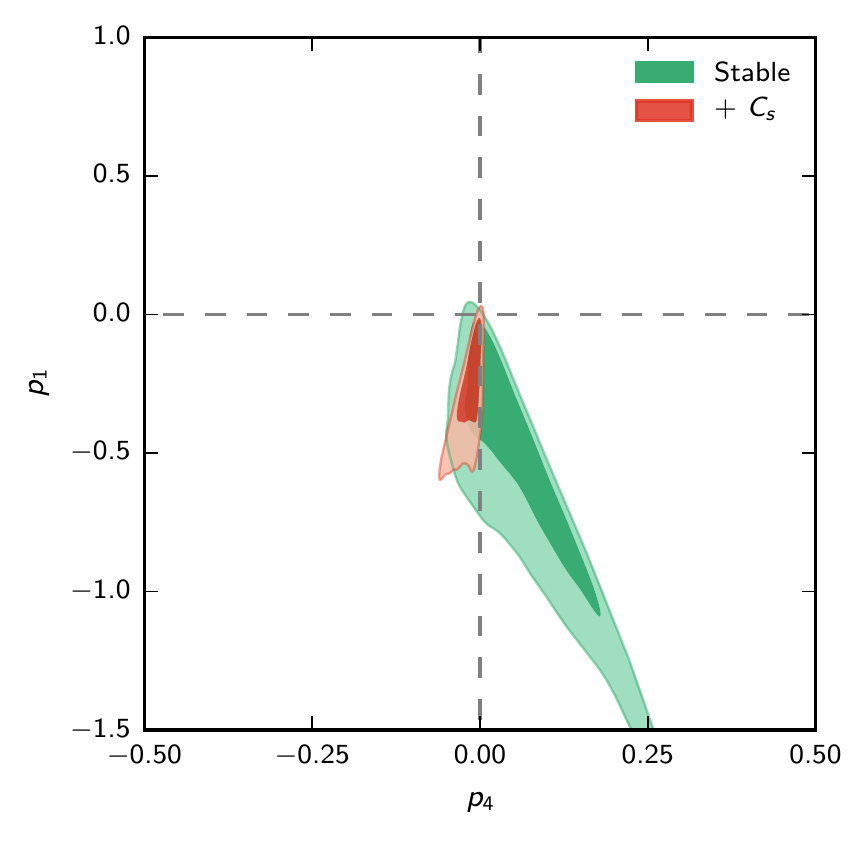}
\includegraphics[scale=0.58]{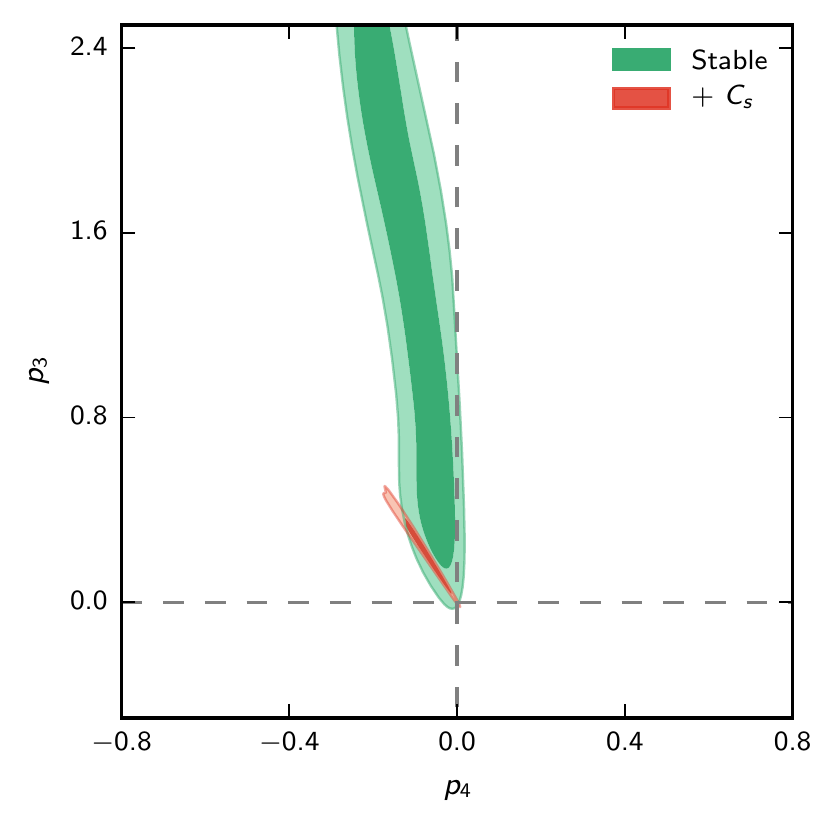}
\caption{\label{fig:viability_regions} 
We illustrate here the role of the viability conditions in shaping the EFT parameter constraints, for two possible scenarios: stable (green), stable \& scalar subluminality (red). Top: from left to right, viability regions for $p_4=0$, $p_3=0$, $p_1=0$, and where we have fixed $\Omega_m^0 =0.27$. Bottom: two-dimensional posteriors for the corresponding pairs of EFT parameters when considering the PLANCK dataset. We marginalize over the cosmological parameters. For visualisation clarity, the EFT parameters not shown in each plot are set, in turn, to zero. The viability regions  are thus different from those of the 3D case presented below in Fig.~\ref{fig:2Dconstraints-ordzero}. Viability conditions tightly reduce the width of the parameter space, data further reduce the allowed regions. Interestingly the $\Lambda$CDM case is always at the corner of the viable space.
}  
\end{figure*}

\subsection{The role of viability conditions}

\begin{figure*}
\centering
\includegraphics[scale=0.42]{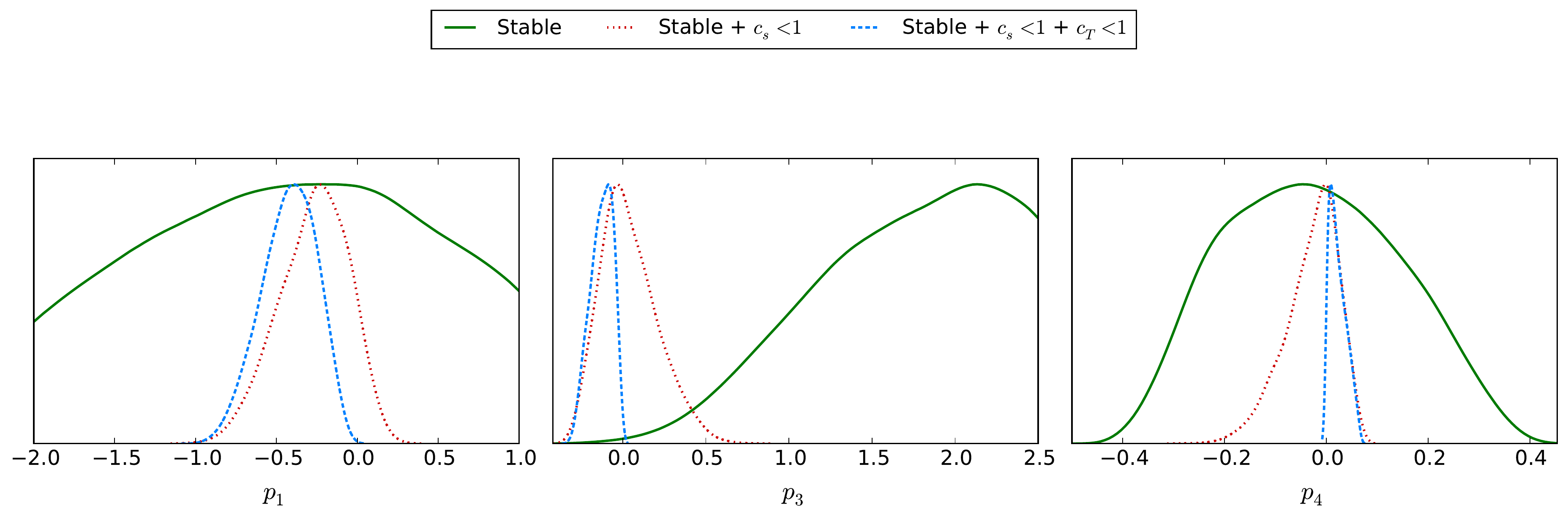}
\caption{\label{fig:posteriors-ordzero} Planck constraints on the {\it 3D-}model,  the minimal EFT  extension of the six-parameter $\Lambda$CDM model. The marginalised,  posterior distribution for $p_1$, $p_3$ and $p_4$  is shown. The likelihood analysis is  carried out by also letting vary the six parameters of the standard $\Lambda$CDM model and the foreground parameters. The constraining power of theoretical priors  is also shown: The solid line shows the posterior obtained after removing  the portion of the parameter space in which unstable theories live. Likelihood intervals obtained by requiring that scalar perturbations, in stable theories,  propagate at subliminal speed are  shown by  dotted lines. Dashed lines show results when  also tensor perturbations are forced to propagate at subluminal speed.}  
\end{figure*}

To begin, let us emphasise one of our main results: the main role of the theoretical  viability conditions in determining the parameter constraints. 
In Fig. \ref{fig:viability_regions} we plot the two-dimensional posterior PDF of the {\it model 3D} (bottom panels) and we compare it with the areas delimited purely by the viability requirements (top panels), \emph{i.e.}, without data. To make the relation between regions of viability and posteriors clearer, here we do not marginalize over the third EFT parameter but only on the six $\Lambda$CDM ones, fixing in turn one of the three $p_i$ parameters to zero. These plots show, on the one hand, the important role of viability conditions in shaping the posterior distributions. Another important feature highlighted here is that within the space of theories considered, and with the additional constraint of reproducing the same expansion history as $\Lambda$CDM, setting all couplings to zero lies at the edge, more precisely on a tight corner, of the allowed parameter space. (see also Figs 1 and 2 of Ref.~\cite{pheno} on this). In other words, $\Lambda$CDM is an extremal among all modified gravity models with the same equation of state $w = -1$. This implies that, when we sample the theory space by means of the Markov chain algorithm, the chance of hitting $\Lambda$CDM model is extremely small.  In practice, the standard $\Lambda$CDM model is never reached by the chain.   What typically happens is that the $\chi$-squared minimisation tends to go, from a given stable point, towards the origin of the parameter space ($\Lambda$CDM). However, while approaching $\Lambda$CDM, the allowed region becomes a tight throat, and finding stable theories becomes more and more difficult. As a result, the posteriors are often decentered  from $\Lambda$CDM, without necessarily implying a better fit of the data (\emph{i.e.} see Fig.~\ref{fig:posteriors-ordzero}). In this theory landscape, $\Lambda$CDM is truly an extremal among all models with the same background expansion history.

\subsection{The 3D-Model}

\begin{figure*}
\centering
\vspace{0.8cm}
\includegraphics[scale=0.59]{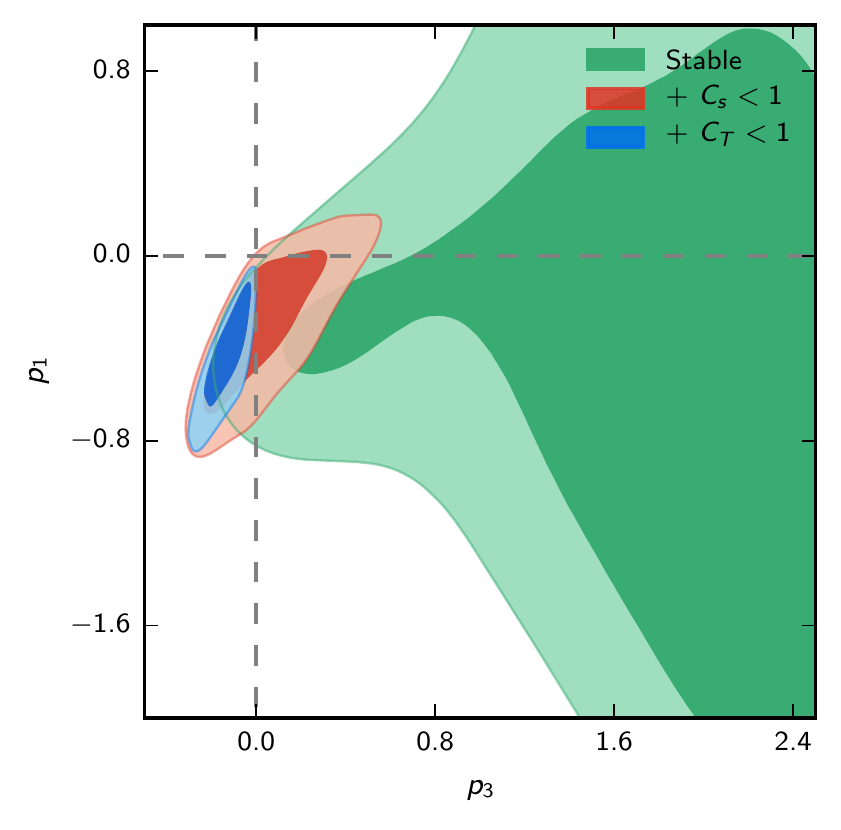}
\includegraphics[scale=0.59]{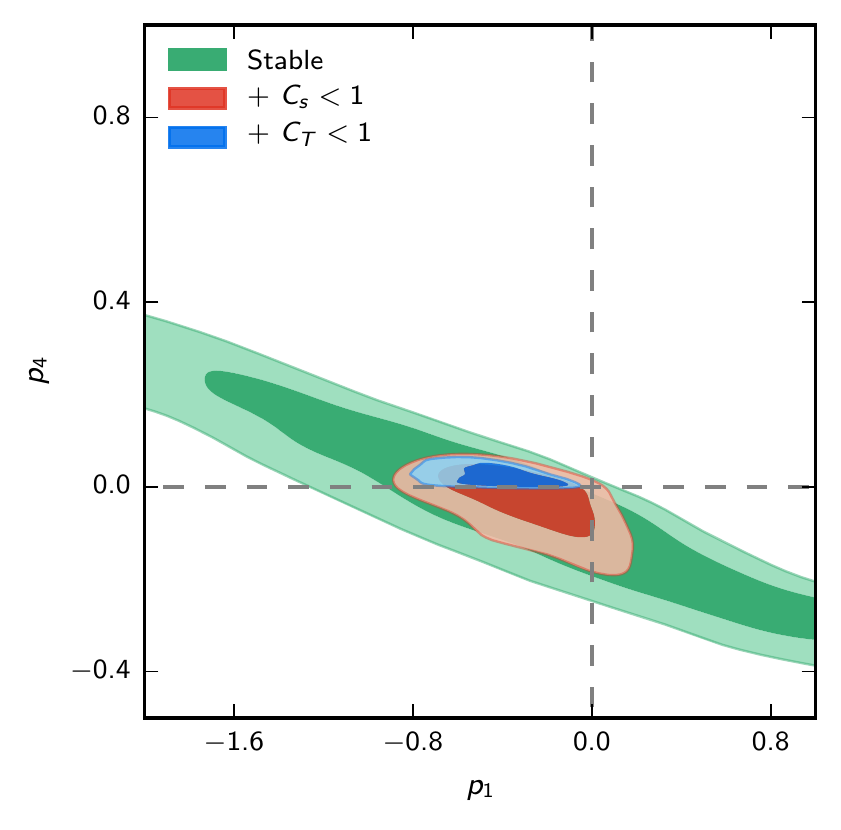}
\includegraphics[scale=0.59]{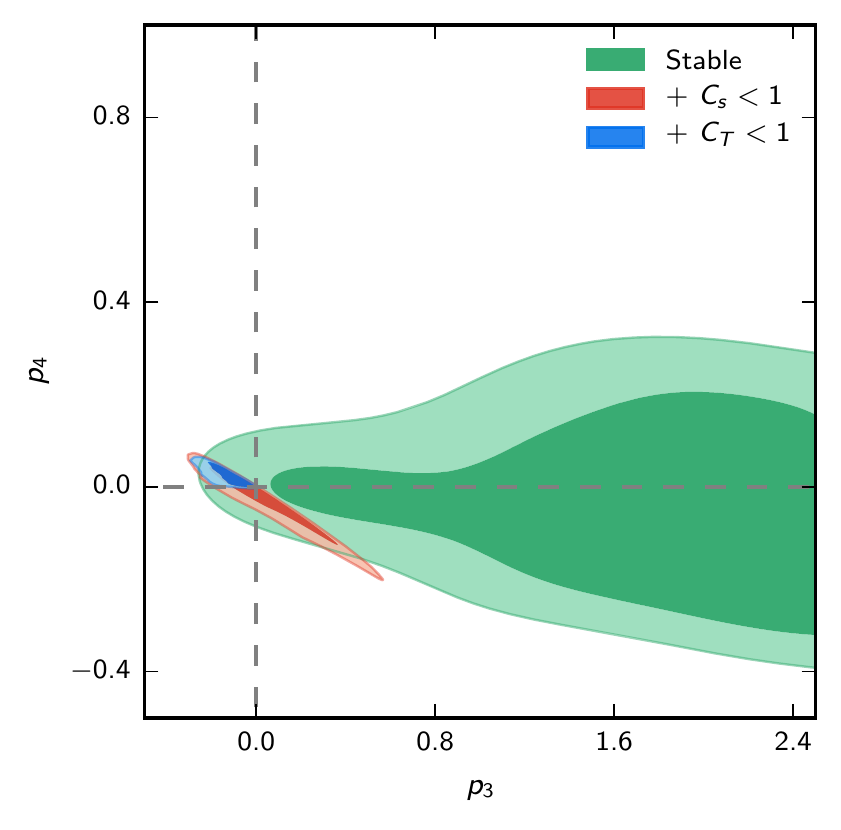}
\caption{\label{fig:2Dconstraints-ordzero} 
Planck constraints on {\it model 3D},  the minimal EFT  extension of the six-parameter $\Lambda$CDM model. The 2D, marginalised,  posterior distributions for the zero$-th$ order EFT parameters $p_1$, $p_3$ and $p_4$, are shown. Marginalisation is over the remaining EFT parameters and  the six $\Lambda$CDM parameters shown in Table \ref{Tab1}.
Contours display the  $68\%$ and $95\%$ c.l.. Different colours corresponds to the three different viability scenarios explained in \ref{sec:viabcond}.}  
\end{figure*}

In Fig.~\ref{fig:posteriors-ordzero} and~\ref{fig:2Dconstraints-ordzero}  we present  constraints on the  minimal EFT extension  of the standard model of cosmology (\emph{3D model}). The most prominent feature visible in the one-dimensional posteriors of $p_1$, $p_3$ and $p_4$ in Fig~\ref{fig:posteriors-ordzero}  is  that the theoretical requirement of subluminality (for  both scalar and tensor modes)   significantly narrows the posterior interval of the EFT parameters. The theoretical viability conditions are thus  powerful instruments  that complement and  increase 
the discriminatory power of data.   Specifically,  the main effect of the subluminality prior on  $p_1$ and $p_4$ parameters is a reduction of the distribution width around the null value, while the net effect on  $p_{3}$  is instead  to shift it towards   lower  values,  compatible with the absence of  modified gravity signals. 

The amplitude of the posterior intervals suggest that the best fitting EFT parameters are small, of order 1, thus confirming the reliability of the series expansions \eqref{tay1}-\eqref{tay3} 
in featuring the essential scaling of the coupling functions $\mu(t)$, $\mu_3(t)$ and $\epsilon_4(t)$.
While the one dimensional PDF peaks at around zero for both $p_3$ and $p_4$,  a negative value of $p_1$ is preferred under any viability  conditions. 
This  would apparently imply that the  $C_{l}$ data are best fitted by non-standard gravity models predicting larger temperature fluctuations than $\Lambda$CDM at low $l$ (see Fig~\ref{fig:spectra}). This fact seems thus at variance with  observational lack of power at low $l$ multipoles in the CMB TT spectrum. The paradox is easily solved by 
inspecting  the 2-dimensional posteriors of the parameters. Indeed,  Fig.~\ref{fig:2Dconstraints-ordzero} clearly shows that the EFT parameter $p_1$ is strongly degenerate  with $p_4$.
The observed  anti-correlation results from the fact that both 
excite  the same  range of multipoles of the temperature power spectrum (see Fig~\ref{fig:spectra}). In other 
terms, an increase in power generated by a negative $p_1$ is mostly compensated by the suppression mechanism activated by a positive value of  $p_4$. 

The bi-dimensional projected posterior PDF of the {\it model 3D} parameters is shown  in Fig. \ref{fig:2Dconstraints-ordzero} for the three combinations of parameters. Here, when displaying two EFT parameters, we are marginalizing over other parameters. This explains the differences between Fig~\ref{fig:viability_regions} and Fig~ \ref{fig:2Dconstraints-ordzero}.   This picture shows that asking for subluminal velocities  considerably narrows the confidence regions.  Interestingly, the viability  priors not only  impose tighter constraints as compared to those derived from cosmological measurements alone, but they also compensate  both the statistical insensitivity and the parameter degeneracy. 

As for the goodness of the fit and its implications for models selection,  Table \ref{Tab1}  shows that the observed decreasing in the $\chi^2$ value associated to the best fitting EFT model  is not significant, {\it i.e}  it is not enough for  the best fitting {\it model 3D}, which has $3$  more degrees of freedom  than the standard model of cosmology,   to be statistically preferred.  Additionally, Table \ref{Tab1} shows that  the best fitting value of the $\Lambda$CDM parameters are close, and statistically indistinguishable, from those of the reference $\Lambda$CDM model calibrated by Planck.  

One might expect MG models to possibly improve the CMB fit, by allowing a power suppression in the temperature low-multipoles spectrum. The fact that the models here considered do not improve the $\chi^2$ of $\Lambda$CDM, is, however, in agreement with the previous analysis in the EFT framework presented in \cite{Bellini:2015xja}.  Note also that the correlation between the parameters seems to introduce an additional rigidity. For example, the degeneracy between $p_1$ and $p_4$ creates a mechanism of compensation that prevents these parameters to suppress power at low-multipoles. Power suppression can therefore be obtained only for positive values of $p_3$. However, the improvement of the $\chi^2$ of $\sim 2$ is usually related to the ability of a model to suppress power around the anomalous dip at $l \sim 20$ while $p_3$ affects the power spectrum only at scales $l<10$. Additionally, when imposing the scalar subluminality, the allowed magnitude of $p_3$ becomes significantly smaller and positive values of $p_3$ are completely discarded when tensor subluminality is imposed. This explains why an improvement in the $\chi^2$ is obtained only in the "3D-stable" case.

We remark that among the three EFT parameters, $p_3$ is the least constrained. Linear perturbation equations  which describe the growth of the matter distributions in the universe are weakly sensitive to this parameter that controls the emergence of galileon/Horndeski-3 type models in our modified gravity landscape.  This result is the consequence of  what we saw in  Fig~\ref{fig:spectra},  that is,  
the amplitude of the linear  power spectrum of temperature fluctuations are mostly sensitive to $p_3$ only on very large cosmic scales (low $l$),  where CMB data are fewer and affected by large statistical and systematic noise. 

\begin{figure*}
\includegraphics[scale=0.35]{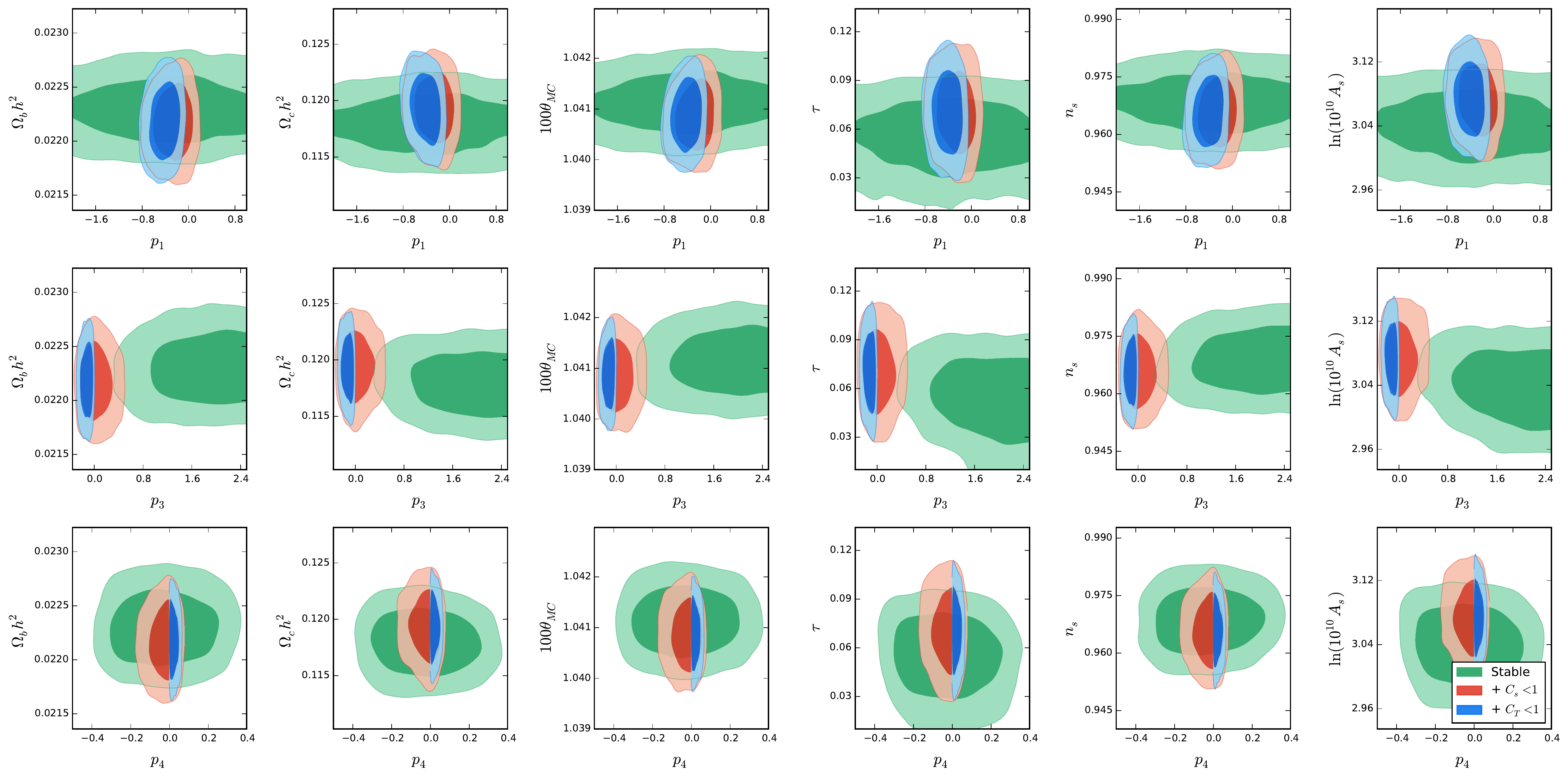}
\caption{\label{fig:ordzero_deg} 
The bidimensional, marginalised, posterior PDF projected onto various planes defined by the zero$-th$ order EFT parameters $p_1$, $p_3$ $p_4$,  and the 6 parameters of the 
Standard $\Lambda$CDM model is shown.}
\end{figure*}

\subsection{EFT and the other (standard) parameters of $\Lambda$CDM}

Fig. \ref{fig:ordzero_deg}  shows that, contrary to  naive expectations,  adding extra dimensions to the dark energy parameter space does not shift nor enlarge signficantly the confidence interval of the six $\Lambda$CDM parameters measured  by Planck. This is due to the fact that the $p_i$ parameters do not show appreciable degeneracy with the $\Lambda$CDM parameters. In other terms, CMB measurements of the 6-parameters of the $\Lambda$CDM model are robust against the inclusion of external couplings controlling the perturbation sector. Our formalism, which displays a clear separation between expansion rate and perturbation sectors, allows one to check this very clearly. The 6 parameters of the $\Lambda$CDM model are essentially insensitive to small variations in the EFT parameters, in the sense that the centers of the Planck error bars are not offset, nor the interval of confidence degraded (Tab. \ref{Tab1}). 
Interestingly, the insensitivity to small variations in the  EFT parameters is not an  harmful issue. Indeed, as Fig. \ref{fig:ordzero_deg} shows, physical  priors on the viability of alternative gravity models help in beating down the lack of resolution on EFT parameters, thus reducing the uncertainty  associated to their estimate.  

\begin{figure*}
\centering
\includegraphics[scale=0.5]{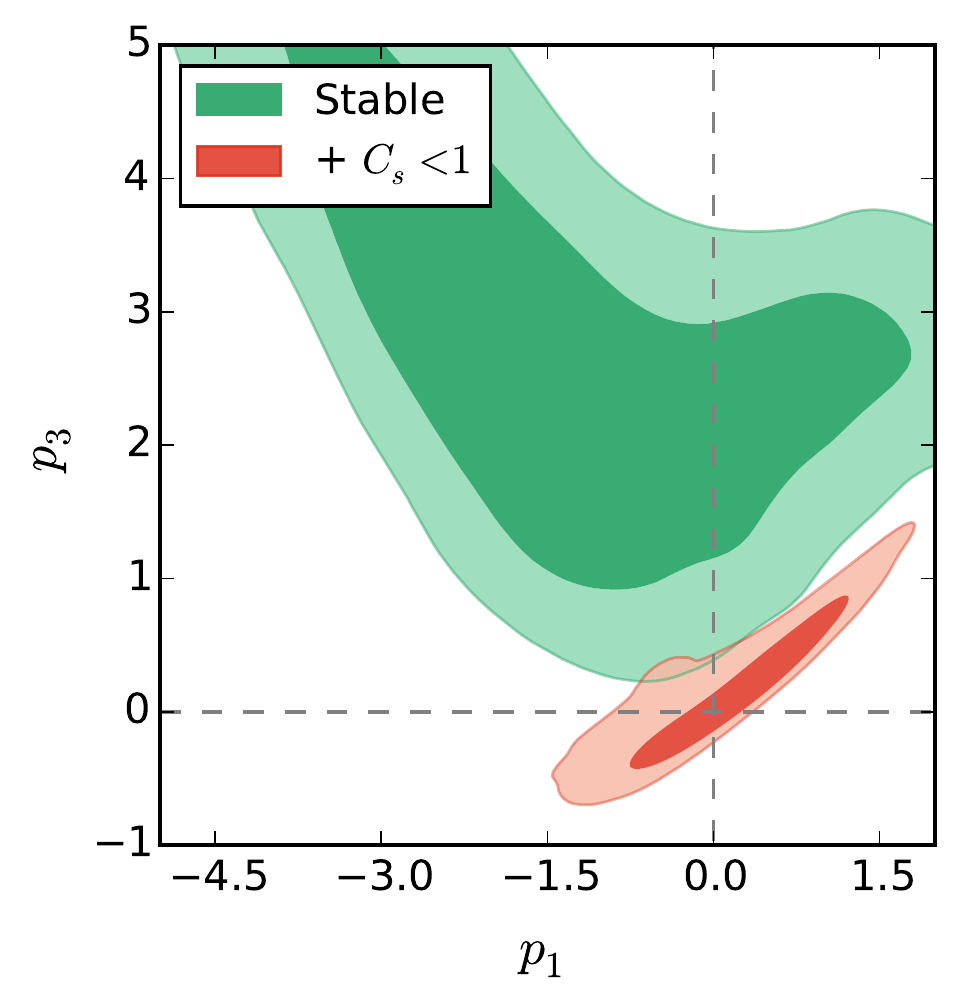}
\includegraphics[scale=0.5]{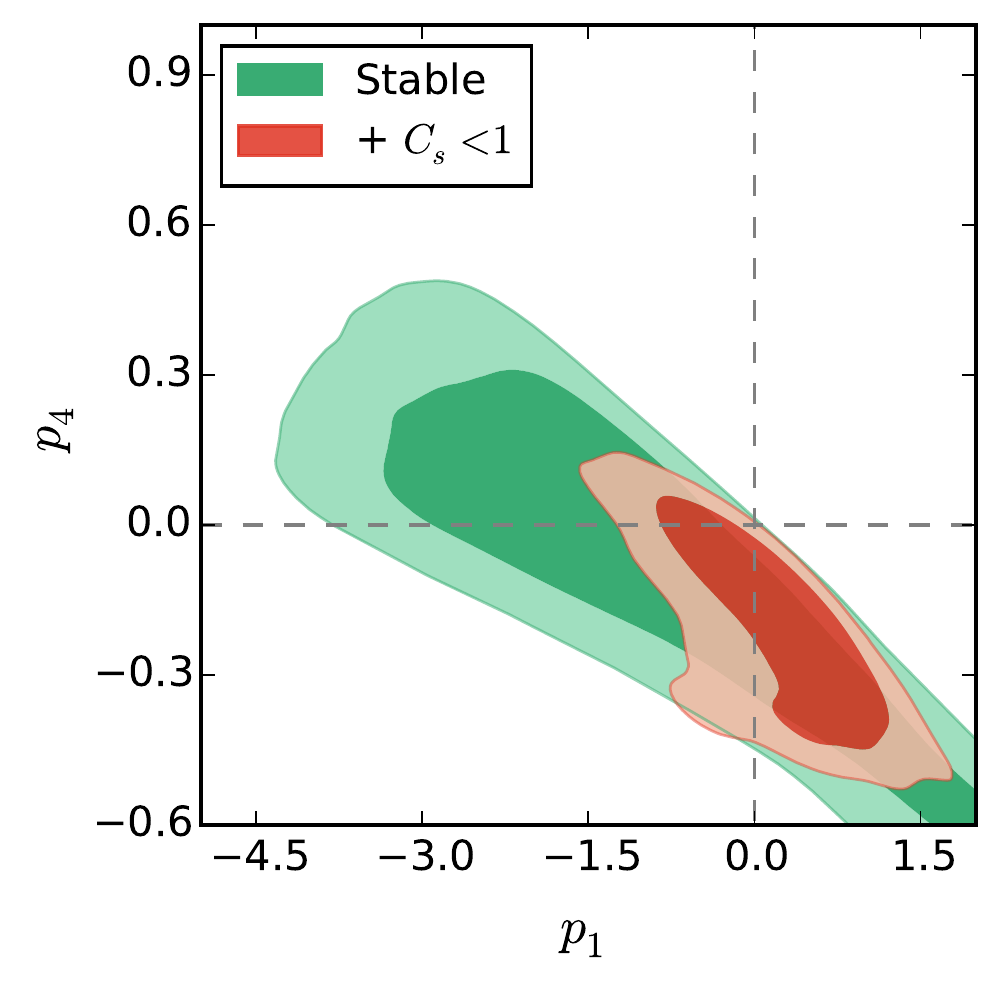}
\includegraphics[scale=0.5]{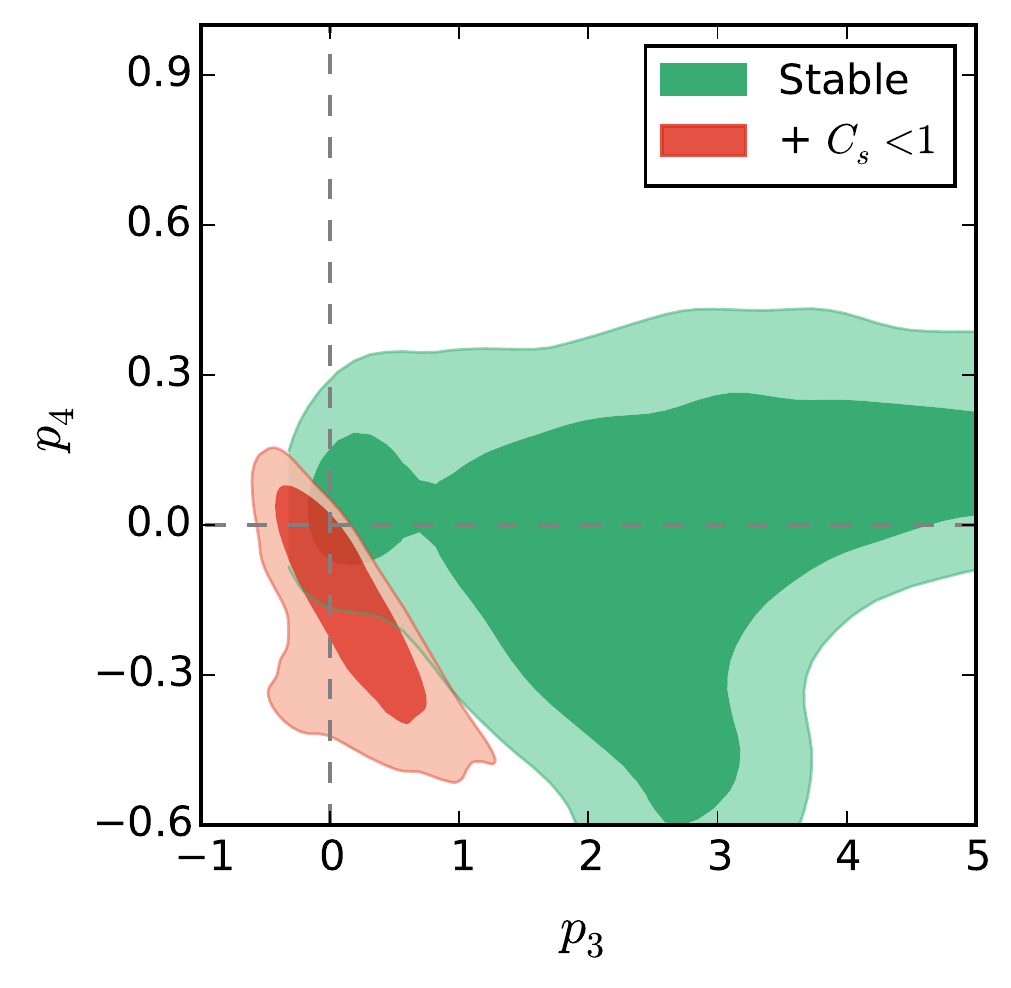}
\includegraphics[scale=0.5]{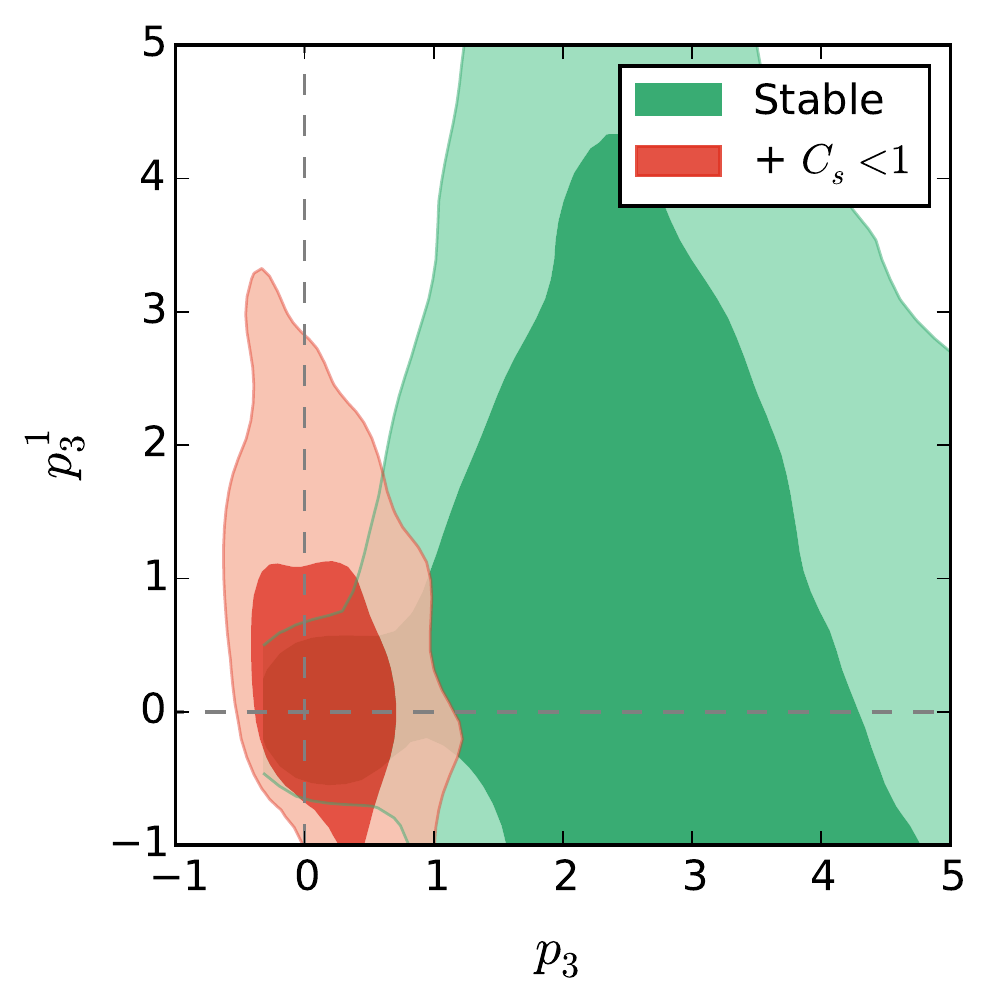}
\includegraphics[scale=0.5]{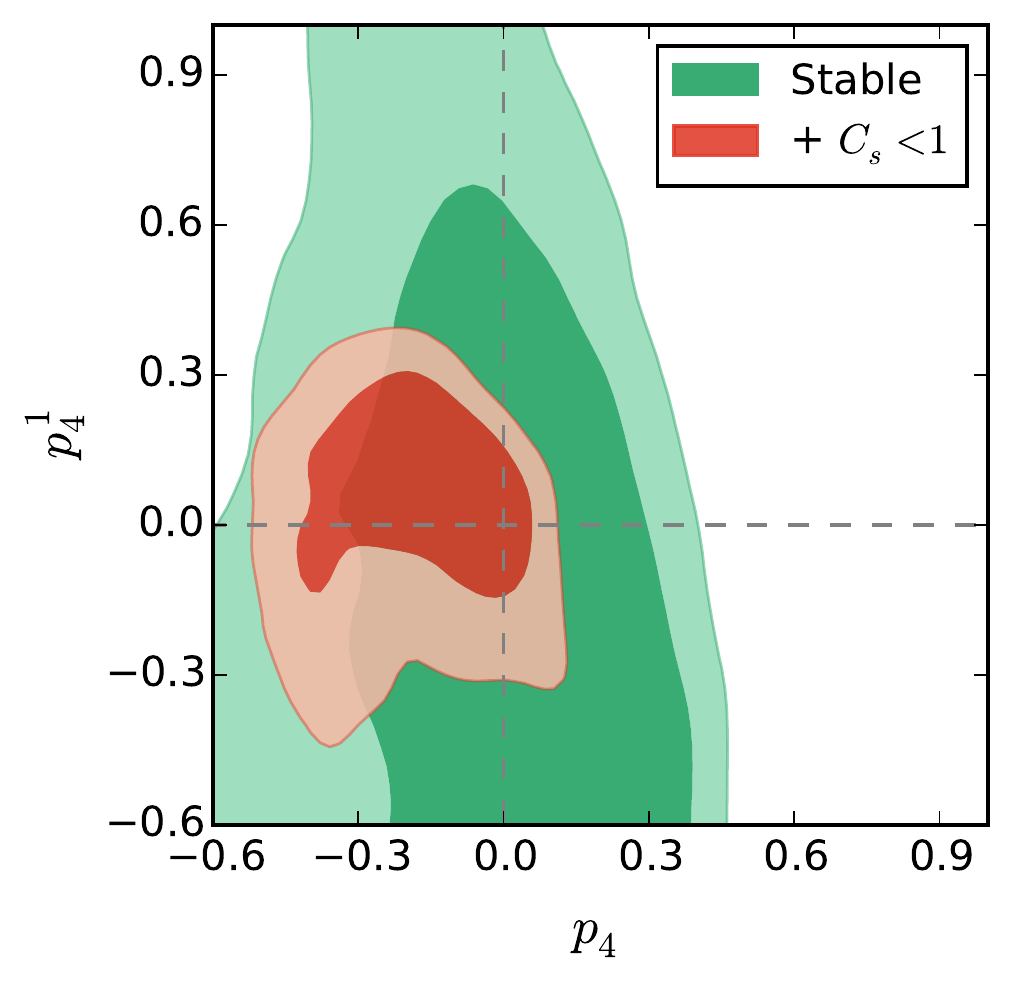}
\caption{\label{fig:2Dconstraints-ord1} 
Planck constraints on the {\it 5D} model,  the maximal EFT  extension of the six-parameter $\Lambda$CDM model explored in this paper. The 2D, marginalised,  posterior PDF 
 for  all the free fitting parameters in Eqs. \eqref{tay1}-\eqref{tay3}   is shown. Marginalization is over the remaining EFT parameters of model $\it 5D$ and also over  the six $\Lambda$CDM parameters shown in Tab. \ref{Tab1}. The likelihood contours display the  $68\%$ and $95\%$ c.l.. In these plots the blue contours are not displayed because the Markov chain was virtually unable to find theories satisfying \emph{all} viability conditions. }  
\end{figure*}
\subsection{The 5D model}

The absence of any  $3$-parameters EFT model performing  better than $\Lambda$CDM, together with the above remarks on the statistical power of the viability constraints, suggest
to extend our parameter space by including two additional degrees of freedom, represented by the terms   $\pu_3$ and $\pu_4$ in Eqs. \eqref{tay2}  and \eqref{tay3}. By this choice, the  parameterization  becomes more flexible and,  in  principle, able to capture subtler time behaviors of the EFT coupling functions. This, in turn, should allow one  to 
explore the space of  Horndeski theories in a finer way, and eventually  single out models, if any,  that outperform  the standard one.
Results of this likelihood analysis  are displayed in Fig.~\ref{fig:2Dconstraints-ord1} and  in the last column of Table \ref{Tab1}. 

As expected, EFT parameters are somewhat less constrained when the \emph{5D} model is considered, since we are dealing with more degrees of freedom. Nevertheless, the contours in the first row of Fig.~\ref{fig:2Dconstraints-ord1} show that the degeneracies between the $0^{th}$-order parameters are unaffected by the additional parameters. Also, no remarkable degeneracies are evident between the $0^{th}$-order parameters $p_3$, $p_4$ and the $1^{st}$-order parameters $\pu_3$, $\pu_4$. This implies that the \emph{3D} model already catches the main features of the modifications of gravity. However, the enlargement of the constraints completely washes out the indication of a preferred negative $p_1$. As a consequence,  no signals of new physics emerge, also within this enlarged parameter space. 

{Finally, we note that when considering the \emph{5D} model, while a stable region  corresponding at least to the case ${p_3^{(1)}, p_4^{(1)}} = 0$ exists, catching these few stable points in the full parameter space becomes very improbable, due to the increased volume effect. Practically, these stable regions are inaccessible to the MCMC chain---this is the reason why no blue contour  is displayed in Fig.~\ref{fig:2Dconstraints-ord1}.
Since such a difficulty is a volume effect (\emph{i.e.}, roughly, the smallness of the quantity \emph{volume of stable theories}/\emph{total volume}), it is not surprising that it can be made worse by going to higher dimensions in the theory space. This further shows the difficulty of finding stable theories around $\Lambda$CDM.

\begin{table*} 
\scriptsize
\begin{center}
\begin{tabular}{|l|| c| c| c| c| c| }
\hline
 & \multicolumn{1}{c}{{\bf $\Lambda$CDM}} & \multicolumn{3}{|c|}{{\bf 3D}} & \multicolumn{1}{|c|}{{\bf 5D}} \\ \hline
 Parameters &  & Stable & Stable $ \;$ \& $\,$ $c_s\! <\!1$ & Stable $\;\&\;$  $c_s\! <\!1$ $\; \& \;$  $c_T\! <\!1$  & Stable $ \;$ \& $\,$ $c_s\! <\!1$  \\ \hline
$\Omega_{\rm b} h^2$  &  $0.02224 \pm 0.00023$ &  $0.02231 \pm  0.00024$ &  $0.02217 \pm  0.00022$ & $ 0.02219 \pm 0.00023 $ & $0.02215 \pm 0.00023$  \\  \hline
$\Omega_{\rm c} h^2$  &    $0.1186 \pm 0.0020$ & $0.1180 \pm 0.0021$ & $0.1194  \pm 0.0019$ & $0.1194 \pm 0.0019$ & $ 0.1198\pm 0.0020$  \\ \hline
$100\theta$   &     $1.04101 \pm 0.00047$ &  $1.04113 \pm 0.00049$ &  $1.04087 \pm 0.00044$ &  $1.04089 \pm 0.00047$ & $1.04084 \pm 0.00047$  \\ \hline
$\tau$ &    $0.066 \pm 0.017$  & $0.060 \pm 0.020$ &  $0.075 \pm  0.015$ & $0.073 \pm  0.016$ & $0.073 \pm 0.016$  \\ \hline
$n_{\rm s}$ &  $0.9675 \pm 0.0060$  & $ 0.9687 \pm 0.0061  $ &  $0.9656 \pm  0.0058$ & $ 0.9657 \pm 0.0059$ & $0.9650 \pm 0.0059$  \\ \hline
$\log(10^{10} A_{\rm s})$  &  $3.062 \pm 0.030$    & $3.050 \pm 0.039 $ &  $3.082 \pm  0.027$ & $3.078 \pm 0.029$ & $3.080 \pm 0.029$ \\ \hline
$p_1$   &   $ - $  & $-0.43^{+0.95}_{-0.21}$ & $-0.28^{+0.17}_{-0.20}$& $-0.42_{-0.17}^{+0.21}$ & $0.10^{+0.58}_{-0.37}$ \\ \hline
$p_3$  &   $ -$    & $>0.13$ (95\% c.l.) &  $0.04 \pm  0.17$ & $-0.12_{-0.06}^{+0.08}$&  $0.13^{+0.28}_{-0.40}$ \\ \hline
$p_4$  &   $- $    & $-0.03_{-0.19}^{+0.16}$ & $-0.030^{+0.068}_{-0.035}$& $0.023_{-0.023}^{+0.009}$ & $-0.18^{+0.28}_{-0.13}$ \\ \hline
$p_3^1$ &$- $ &$- $ &$- $ & $- $& $0.41^{+0.39}_{-0.91}$ \\ \hline
$p_4^1$ & $- $& $- $&$- $ & $- $& $0.03^{+0.18}_{-0.11}$ \\ \hline 
\hline
$\chi^2$ &  11276.97 & 11276.22 & 11278.46 & 11278.62 & 11277.52  \\ \hline
\end{tabular}
\caption{\footnotesize Constraints on the parameters for the \emph{3D} and \emph{5D} models, together with the six standard $\Lambda$CDM parameters (the notation for the latter is the standard one, see \emph{e.g.}~\cite{Ade:2015cp}).}
\label{Tab1}
\end{center}
\end{table*}

\section{Results: Constraints on cosmological observables} \label{sec:resultsOBS}
A complementary approach consists in constraining not  the space of theories, as we did above, but quantities that are closer to direct cosmological observables. 
To this purpose, we consider the functions $\mu_{\rm MG} (t)$ (the effective Newton constant), $\gamma_{\rm MG}(t)$ (the gravitational slip) and $\Sigma (t)$ (the lensing potential), as defined in \eqref{pi} and \eqref{Sigma} . Indeed, these are the cosmological functions that are directly constrained by most observational probes. Any deviations from the unity of these functions, at any redshift, is notably considered a smoking gun for modified gravity. 

The main results of our analysis are presented in Fig.~\ref{fig:mu-gamma} and show the importance of not neglecting viability conditions in phenomenological constraints.

Firstly we consider the $\mu_{\rm MG} (t)$-$\gamma_{\rm MG}(t)$ plane (left part of  Fig.~\ref{fig:mu-gamma}) that is where most of the recent analysis of the Planck Collaboration has been focused ~\cite{Ade:2015mg}. Interestingly, the Planck Collaboration highlighted that the combination of CMB and redshift-space distortions (RSD) and/or galaxy weak-lensing data (WL) indicate a deviation from the standard value, at about 3 sigma, of both these functions at redshift $z=0$ (see Fig.~14 of Ref.~\cite{Ade:2015mg}). Namely, values of $\mu_{\rm MG}(0)$ lower than 1 and values of $\gamma_{\rm MG}(0)$ higher than 1 seem to be preferred by this combination of probes. This result is largely driven by the preference of these probes for a $rms$ density fluctuations on the scale of $8 h^{-1}$ Mpc lower than expected in standard gravity \cite{Heymans:2013fya, Beutler:2013yhm, Samushia:2013yga, Anderson:2013zyy}. Whether  these are just unidentified systematics or real physical indications, it is worth trying to see if these results are compatible with any healthy theory of the Horndeski class.

The main methodological difference between our approach and that adopted in \cite{Ade:2015mg} is the way in which the time scaling of the function $\mu_{\rm MG}$ and $\gamma_{\rm MG}$ is parameterised. There, two phenomenological behaviours are chosen for these functions: one in which the time variation is proportional to the dark energy density (dark-energy related parameterisation) and the other in which the variation is proportional to $1-a$ (time-related parameterisation). In both cases, $\mu_{\rm MG}$ and $\gamma_{\rm MG}$ evolve independently of one another. Here, the expressions of $\mu_{\rm MG}$ and $\gamma_{\rm MG}$  are theoretically determined within the context of the EFT theory and they can be tracked back to underlying physical healthy theories. Remarkably, such functions do not have a generic behaviour but, rather, they present definite features and specific correlations, especially when all viability conditions are used to select models~\cite{pheno,pheno2}.

\begin{figure*}
\centering
\includegraphics[scale=0.7]{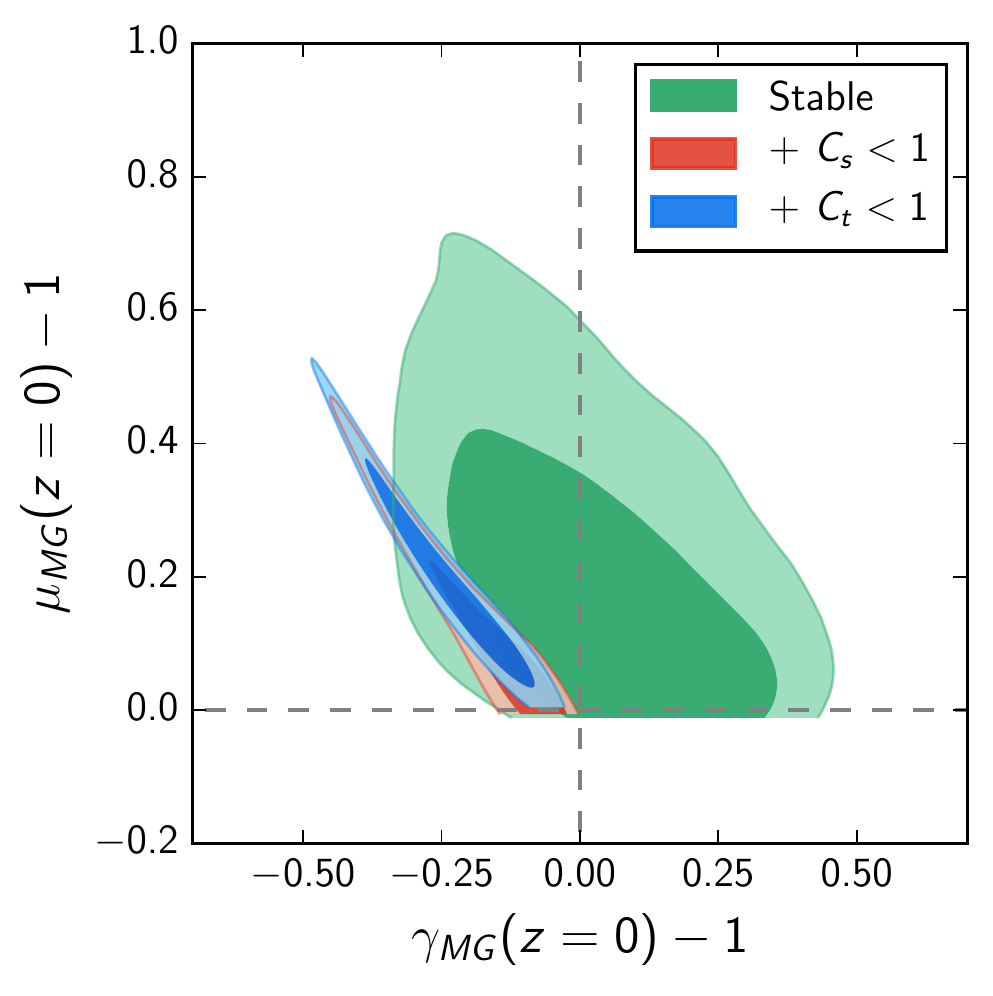}
\vspace{1.5cm}
\includegraphics[scale=0.7]{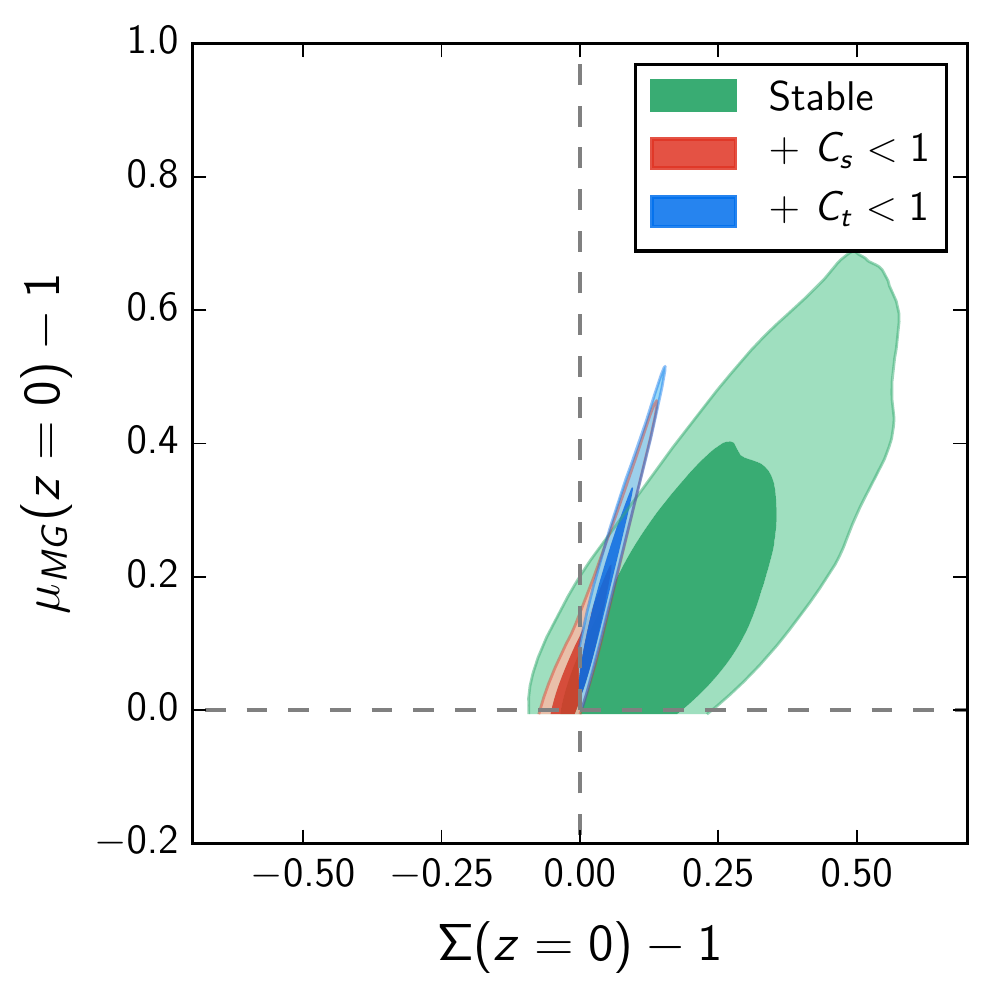}
\caption{\label{fig:mu-gamma} 
 68 \% and 95 \% contour plots for the couple of functions ($\mu_{\rm MG}$, $\gamma_{\rm MG}$) and ($\mu_{\rm MG}$, $\Sigma$) evaluated at present time by using Eqs.~\eqref{geff}, ~\eqref{postn}, ~\eqref{Sigma} and the minimal parameterisation of the {\it 3D} model.}  
\end{figure*}
 
First we note from the results in Fig.~\ref{fig:mu-gamma} that there is no stable EFT model that lives in the portion of the space of observables characterised by a negative value of $\mu_{\rm MG}-1$ today. Therefore a large part of the region allowed by phenomenological models in \cite{Ade:2015mg} is {\it a-priori} excluded as unphysical. More importantly, the best-fitting region arising by combining CMB and RSD and/or WL lays in the unstable region. That means that, even if confirmed by future experiments, that signal would incompatible with the entire class of Horndeski theories. 

Even more stringent conclusions can be drawn if subluminal propagation speed of scalar and tensor perturbations is required. In this case,  we find an additional tight constraint on the present value of $\gamma_{\rm MG}$ that is strictly lower than unity in all viable  Horndeski theories.

The healthy EFT models that satisfy to the most stringent set of our conditions (no ghosts and gradient instabilities, subluminal propagation speed for scalar and  tensor perturbations)  are indeed  characterized by a positive value of $\mu_{\rm MG}-1$ \emph{and} a negative value of $\gamma_{\rm MG}-1$.  As a result, the favoured values of those parameters are pulled back towards the $\Lambda$CDM region, with no  compelling evidence for any better description of data than that offered by the $\Lambda$CDM model.

The fact that $\Lambda$CDM seems to live on the border of the 2$\sigma$ confidence level of the most constrained posterior likelihood (blue region in Fig. \ref{fig:mu-gamma}) is just an artefact due to the sampling behaviour when the parameter space is cut by viability conditions. As a matter of fact, the stability regions for these models is a sharp corner of which the $\Lambda$CDM models occupies the vertex. This regions is thus virtually inaccessible by the Markov Chain.

The one-dimensional posteriors of $\mu_{\rm MG}$ and $\gamma_{\rm MG}$ from PLANCK, when healthy conditions on the theories are imposed, are depicted in Fig \ref{fig:mu-gamma2}.

Despite the larger contours, the same behaviour (value of $\mu_{\rm MG}-1$ close to positive and negative value of $\gamma_{\rm MG}-1$) is confirmed by the 5D-model.

\begin{figure*}
\centering
\includegraphics[scale=0.45]{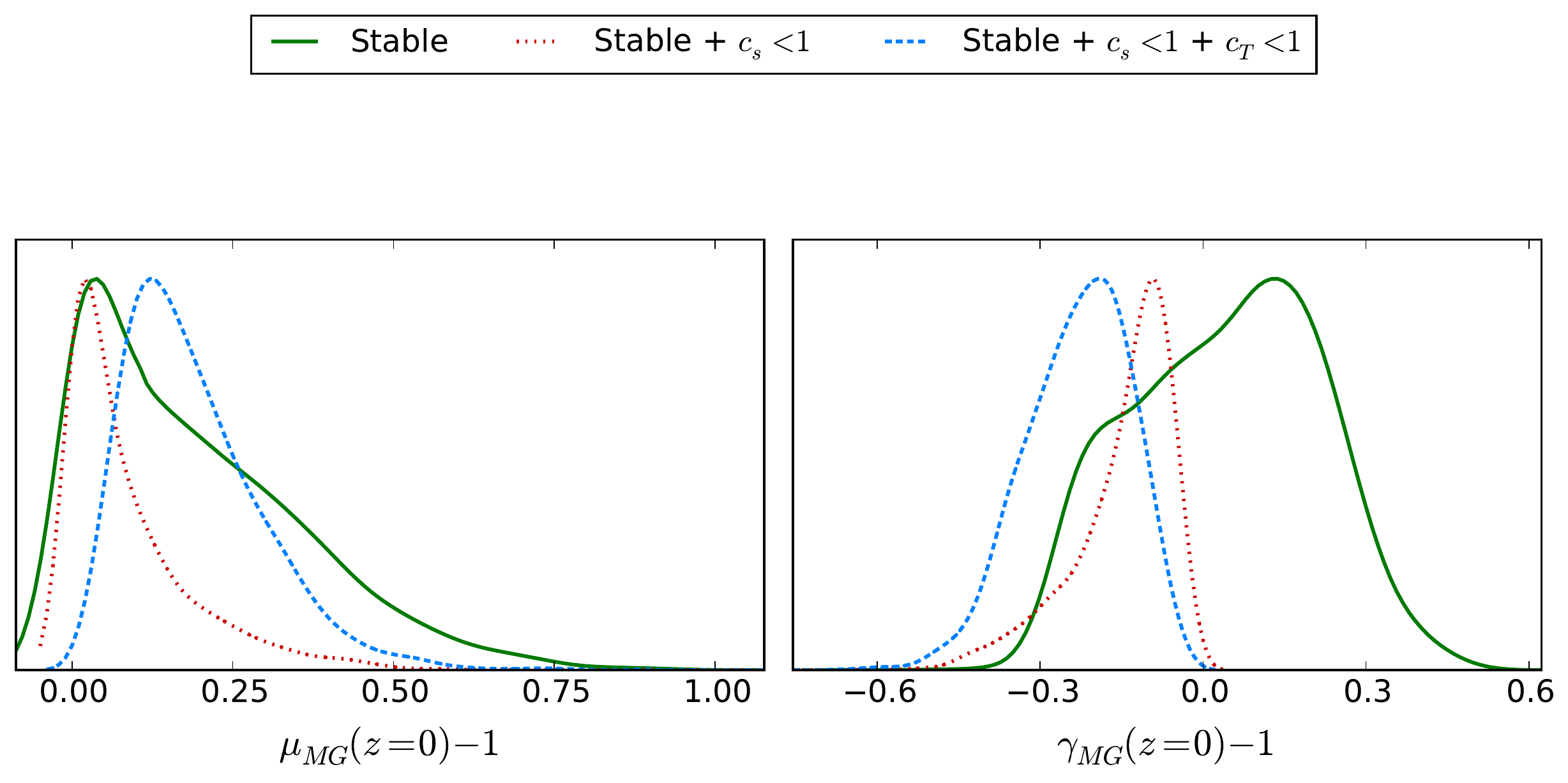}
\caption{\label{fig:mu-gamma2} Case 3D. Posterior distributions for the $\mu_{MG}$ and $\gamma_{MG}$ parameters}  
\end{figure*}

On top of the plane $\mu_{\rm MG}$-$\gamma_{\rm MG} (z=0)$, Fig.~\ref{fig:mu-gamma} also shows results in the $\mu_{\rm MG}$-$\Sigma (z=0)$ plane
as this last quantity is  more  straightforwardly  reconstructed from  weak lensing surveys \cite{Simpson:2012ra, Pogosian:2016pwr}.  As a bonus, by projecting the likelihood onto this plane, we minimise the sampling issues described above.  As a consequence, the fact that current CMB data appears not to be  in tensions with  $\Lambda$CDM predictions stands out even more clearly.   

Interestingly, the more stringent the viability priors imposed to theories,  the more  correlated appear to be the observables $\mu_{MG}$ and $ \Sigma$.
Furthermore,  once the most demanding stability criterium is imposed, i.e. that also tensor modes propagate at subluminal speed, the likelihoods gets confined 
into the first quadrant  of the $\mu_{MG}-\Sigma$ plane. If future and more precise data  were to show that the most likely observables 
$\mu_{MG}$ and $ \Sigma_{\rm MG}$   have opposite sign at $z=0$, i.e.  the likelihood is confined into the second quadrant, then  a definitive  statement about the velocity of gravitational waves could be made ($c_T>1$). Clearly, if the Horndeski class of theory  parameterized in this paper provides the correct interpretation  for  MG gravity signals.

\section {Conclusions} \label{sec:concl}

Exploring beyond the standard model of gravity in the quest for some yet missing physical mechanism that could  explain cosmic acceleration relies on two premises:
high quality astronomical data and flexible parametric scheme that can detect anomalous signals and, at the same time, can interpret them in terms of fundamental physics.
A specific framework  that achieves the latter goal is the  effective field theory of dark energy. In this paper 
we constrain the EFT of DE operators  by means of the  Planck 2015 data.
A key feature of this formalism is that it allows one to analyze the expansion history of the cosmic metric  independently of the perturbation sector.  
We have exploited this possibility to set  the EFT background to  that of a $\Lambda$CDM geometry: as a matter of fact, most  geometrical probes of cosmology  are compatible with the standard $\Lambda$CDM expansion history. 

The simple requirement  that modified gravity theories do not violate fundamental physical  principles such as causality and stability 
results in  stringent  constraints on the accessible regions of the EFT parameter space. Taken together, the  `no-go' regions excluded on theoretical grounds and  
the regions statistically excluded by CMB  data, convincingly suggests that  no scalar-field based extension of GR is more likely than 
the standard gravitational scenario. This result  confirms and strengthens what already found by \cite{Ade:2015mg}  on the basis of 
various purely phenomenological, and thus physically less justified, models.

A complementary line of attack, that gives further angle on the issue, consists of analyzing  the ``observables" $\mu_{\rm MG}$ and $\gamma_{\rm MG}$.
When  their redshift evolution is modelled in a purely phenomenological way, then the ensemble of our cosmological data constrain the local ($z\simeq0$) value of these   quantities to be 
 $\mu_{\rm MG}<1$ and $\gamma_{\rm MG}>1$  (see \emph{e.g.}~\cite{Ade:2015mg}). Physically sound  theories, however, never end up producing such values of $\mu_{\rm MG}$ and $\gamma_{\rm MG}$ \emph{today}. Fig.~\ref{fig:mu-gamma2} shows that  the value of $G_{\rm eff}$ for modified gravity theories not rejected by data, is,   at present epoch  never smaller than the Newton constant.
Additionally,   $\gamma_{\rm MG}(t_0)-1$ becomes negative as soon as the condition for subluminal propagation of perturbations is enforced.

Although we conclude that the models of  dark energy/modified gravity considered here do not seem to be able to outperform predictions of the $\Lambda$CDM model, 
this does not mean that dark energy altogether cannot improve the fit to CMB data. For instance, we could consider scalar field models \emph{beyond} Horndeski~\cite{Gleyzes:2014dya,Gleyzes:2014qga}, or models with more and/or different degrees of freedom than a scalar field. Also, some of our assumptions could be weakened, such as that of imposing no dark energy at early times, which leads to the constraint~\eqref{constraint} among our parameters. The alternative would be that dark energy density does not become subdominant at early epochs. But since we are imposing the $\Lambda$CDM expansion history, this means that its equation of state must mimic non-relativistic matter at early times. Also, along the same ``dark degeneracy" outlined by \cite{Kunz:2007rk,darkdeg}, we could consider models where the physical amount of dark matter today, $\Omega_m^0$, and the geometrical parameter entering the expression of the Hubble rate $H(z)$ as a function of the redshift are two distinct quantities. In the notations of~\cite{pheno,pheno2} this corresponds to considering $\kappa \neq 1$. 

In an upcoming paper, while exploring along some of these directions, we will show results obtained by including in the analysis also observables extracted from low-redshift galaxy datasets. 
 
\acknowledgments
We thank Julien Bel, Jose Beltran, Emilio Bellini, Jason Dossett, Luigi Guzzo, Bin Hu, Stéphane Ilic, Massimiliano Lattanzi, Louis Perenon, Emiliano Sefusatti, Alessandra Silvestri for useful discussions and helpful comments during the completion of this work. We warmly acknowledge the financial support of A*MIDEX project (no ANR-11-IDEX-0001-02) funded by the “Investissements d’Avenir” French Government program, managed by the French National Research Agency (ANR). C.M. is grateful for support from specific project funding of the Labex OCEVU.

\newpage
\appendix

\section{More formulas of the EFT formalism}

Our couplings are implicitly defined by the following action written in unitary gauge. 
\be \label{action}
\begin{split}
S \ =\  & \ S_m[g_{\mu \nu}, \Psi_i] \ +\ \int \! d^4x \, \sqrt{-g} \, \frac{\Mb^2(t)}{2} \, \\[1.2mm]
& \Big[R \, -\,  2 \Lb(t) \, - \, 2 \cb(t) g^{00} \, + \, \mu_2^2(t) (\delta g^{00})^2\, -\, \mu_3(t) \, \delta K \delta g^{00}   + \,  \Big. \\[1.2mm]
&   \epsilon_4(t) \left(\delta K^\mu_{ \ \nu} \, \delta K^\nu_{ \ \mu} -  \delta K^2  +  \frac{\R\,   \delta g^{00}}{2} \right)  \Big] \; \\
\end{split}
\ee
The above describes the background and the first order perturbation equations of the entire set of Horndeski
theories.
Beside the coupling functions, in the action there appears derived quantities, ${\cal C}$ and $\lambda$, that can be identified roughly with the kinetic and potential energy density of dark energy. They can be calculated by applying the background Einstein equations~\cite{EFTOr}:
\begin{align} 
\cb \ &  =   \   \frac{1}{2} ( H \mu - \dot{\mu} - \mu^2 )  - \dot H -  \frac{\rho_m}{2 M^2} \label{c2} \;, \\
\Lb \, &  = \ \frac12 (5 H \mu + \dot \mu + \mu^2  ) + 3 H^2 + \dot H -  \frac{\rho_m}{2 M^2}     \;.
\end{align}

\begin{align} \label{geff}
\mu_{\rm MG}(t) \ &= \ \frac{ M^2(t_0)[1+\epsilon_4(t_0)]^2 }{ M^2(1+\epsilon_4)^2} \ \frac{2 {\cb}  +  \mathring{\mu}_3  - 2 \dot H \epsilon_4 + 2 H \mathring{\epsilon}_ 4 + 2 (\mu + \mathring{\epsilon}_4)^2 \ }{\ 2 {\cb} + \mathring{\mu}_3  - 2 \dot H \epsilon_4 + 2 H \mathring{\epsilon}_ 4 + 2 \dfrac{(\mu + \mathring{\epsilon}_ 4) (\mu - \mu_3)}{1+\epsilon_4} - \dfrac{(\mu - \mu_3)^2}{2 (1+\epsilon_4)^2}  } \ , \\
 \label{postn}
\gs(t) &= 1 - \frac{(\mu + \mathring{\epsilon}_ 4) (\mu + \mu_3 + 2  \mathring{\epsilon}_ 4) - \epsilon_4 (2 \cb +  \mathring{\mu}_3 - 2 \dot H \epsilon_4 + 2 H \mathring{\epsilon}_ 4)}{ 2 \cb +  \mathring{\mu}_3 - 2 \dot H \epsilon_4 + 2 H \mathring{\epsilon}_ 4 + 2 (\mu + \mathring{\epsilon}_ 4)^2}\, ,
\end{align}
where to simplify the notation we have defined with a circle some ``generalized time derivatives", 
\begin{align}
\mathring{\mu}_3  \ &\equiv \ \dot \mu_3 + \mu \mu_3 + H \mu_3 ,\\
\mathring{\epsilon}_4  \ &\equiv \ \dot \epsilon_4 + \mu \epsilon_4 + H \epsilon_4\, .
\end{align}
The function $\Sigma(t)$ can be easily derived from Eqs.~\eqref{geff} and ~\eqref{postn}
\begin{align}
\label{Sigma}
\Sigma(t)&= \frac{\mu_{\rm MG}(t)(1+ \gs(t))}{2} .
\end{align}



\begin{thebibliography}{99}

\bibitem{Ade:2015cp}
  P.~A.~R.~Ade {\it et al.} [Planck Collaboration],
  ``Planck 2015 results. XIII. Cosmological parameters,''
  arXiv:1502.01589 [astro-ph.CO].
  
  \bibitem{Betoule:2014frx} 
  M.~Betoule {\it et al.} [SDSS Collaboration],
  ``Improved cosmological constraints from a joint analysis of the SDSS-II and SNLS supernova samples,''
  Astron.\ Astrophys.\  {\bf 568}, A22 (2014)
    [arXiv:1401.4064 [astro-ph.CO]].
  
  
\bibitem{Bel:2014awa} 
  J.~Bel, P.~Brax, C.~Marinoni and P.~Valageas,
  ``Cosmological tests of modified gravity: constraints on $F(R)$ theories from the galaxy clustering ratio,''
  Phys.\ Rev.\ D {\bf 91}, no. 10, 103503 (2015)
    [arXiv:1406.3347 [astro-ph.CO]].
     
  
\bibitem{Vikhlinin:2008ym}
  A.~Vikhlinin {\it et al.},
  ``Chandra Cluster Cosmology Project III: Cosmological Parameter Constraints,''
  Astrophys.\ J.\  {\bf 692} (2009) 1060
    [arXiv:0812.2720 [astro-ph]].

\bibitem{Ade:2015fva} 
  P.~A.~R.~Ade {\it et al.} [Planck Collaboration],
  ``Planck 2015 results. XXIV. Cosmology from Sunyaev-Zeldovich cluster counts,''
  arXiv:1502.01597 [astro-ph.CO].
  
  \bibitem{Ilic:2015rna} 
  S.~Ilic, A.~Blanchard and M.~Douspis,
  ``X-ray galaxy clusters abundance and mass temperature scaling,''
  Astron.\ Astrophys.\  {\bf 582}, A79 (2015)
    [arXiv:1510.02518 [astro-ph.CO]].
  
\bibitem{Heymans:2013fya}
  C.~Heymans {\it et al.},
  ``CFHTLenS tomographic weak lensing cosmological parameter constraints: Mitigating the impact of intrinsic galaxy alignments,''
  Mon.\ Not.\ Roy.\ Astron.\ Soc.\  {\bf 432} (2013) 2433
    [arXiv:1303.1808 [astro-ph.CO]].
  
\bibitem{Battye:2013xqa}
  R.~A.~Battye and A.~Moss,
  ``Evidence for Massive Neutrinos from Cosmic Microwave Background and Lensing Observations,''
  Phys.\ Rev.\ Lett.\  {\bf 112} (2014) 5,  051303
    [arXiv:1308.5870 [astro-ph.CO]].
    
\bibitem{Raveri:2015maa}
  M.~Raveri,
  ``Is there concordance within the concordance $\Lambda$CDM model?,''
  arXiv:1510.00688 [astro-ph.CO].

  \bibitem{Macaulay2013}
  E.~Macaulay, I.~K.~Wehus and H.~K.~Eriksen,
  ``Lower Growth Rate from Recent Redshift Space Distortion Measurements than Expected from Planck,''
  Phys.\ Rev.\ Lett.\  {\bf 111}, no. 16, 161301 (2013)
  [arXiv:1303.6583 [astro-ph.CO]].
   

\bibitem{delaTorre:2013rpa} 
  S.~de la Torre {\it et al.},
  ``The VIMOS Public Extragalactic Redshift Survey (VIPERS). Galaxy clustering and redshift-space distortions at z=0.8 in the first data release,''
  Astron.\ Astrophys.\  {\bf 557}, A54 (2013)
    [arXiv:1303.2622 [astro-ph.CO]].

\bibitem{Beutler:2013yhm}
  F.~Beutler {\it et al.} [BOSS Collaboration],
  Mon.\ Not.\ Roy.\ Astron.\ Soc.\  {\bf 443} (2014) 2,  1065
    [arXiv:1312.4611 [astro-ph.CO]].
  
\bibitem{Samushia:2013yga}
  L.~Samushia {\it et al.},
  ``The clustering of galaxies in the SDSS-III Baryon Oscillation Spectroscopic Survey: measuring growth rate and geometry with anisotropic clustering,''
  Mon.\ Not.\ Roy.\ Astron.\ Soc.\  {\bf 439} (2014) 4,  3504
    [arXiv:1312.4899 [astro-ph.CO]].
  
\bibitem{Anderson:2013zyy}
  L.~Anderson {\it et al.} [BOSS Collaboration],
  ``The clustering of galaxies in the SDSS-III Baryon Oscillation Spectroscopic Survey: baryon acoustic oscillations in the Data Releases 10 and 11 Galaxy samples,''
  Mon.\ Not.\ Roy.\ Astron.\ Soc.\  {\bf 441} (2014) 1,  24
    [arXiv:1312.4877 [astro-ph.CO]].

\bibitem{Steigerwald:2014ava}
  H.~Steigerwald, J.~Bel and C.~Marinoni,
  ``Probing non-standard gravity with the growth index: a background independent analysis,''
  JCAP {\bf 1405} (2014) 042
    [arXiv:1403.0898 [astro-ph.CO]].

\bibitem{Ade:2015mg}
  P.~A.~R.~Ade {\it et al.} [Planck Collaboration],
  ``Planck 2015 results. XIV. Dark energy and modified gravity,''
  arXiv:1502.01590 [astro-ph.CO].
  

\bibitem{DiValentino:2015bja}
  E.~Di Valentino, A.~Melchiorri and J.~Silk,
  ``Cosmological Hints of Modified Gravity ?,''
  arXiv:1509.07501 [astro-ph.CO].
  


  \bibitem{Zhao:2008bn} 
  G.~-B.~Zhao, L.~Pogosian, A.~Silvestri and J.~Zylberberg,
  ``Searching for modified growth patterns with tomographic surveys,''
  Phys.\ Rev.\ D {\bf 79}, 083513 (2009)
  [arXiv:0809.3791 [astro-ph]].
  
   \bibitem{Pogosian:2010tj} 
  L.~Pogosian, A.~Silvestri, K.~Koyama and G.~B.~Zhao,
 ``How to optimally parametrize deviations from General Relativity in the evolution of cosmological perturbations?,''
  Phys.\ Rev.\ D {\bf 81}, 104023 (2010)
  [arXiv:1002.2382 [astro-ph.CO]].
  
\bibitem{Bianchi:2012za}
  D.~Bianchi, L.~Guzzo, E.~Branchini, E.~Majerotto, S.~de la Torre, F.~Marulli, L.~Moscardini and R.~E.~Angulo,
  ``Statistical and systematic errors in redshift-space distortion measurements from large surveys,''
  Mon.\ Not.\ Roy.\ Astron.\ Soc.\  {\bf 427} (2012) 2420
    [arXiv:1203.1545 [astro-ph.CO]].

  
\bibitem{Kitching:2016hvn}
  T.~D.~Kitching, L.~Verde, A.~F.~Heavens and R.~Jimenez,
  ``Discrepancies between CFHTLenS cosmic shear \& Planck: new physics or systematic effects?,''
  arXiv:1602.02960 [astro-ph.CO].
  
\bibitem{Joudaki:2016mvz}
  S.~Joudaki {\it et al.},
  ``CFHTLenS revisited: assessing concordance with Planck including astrophysical systematics,''
  arXiv:1601.05786 [astro-ph.CO].

\bibitem{Euclid}
http://www.euclid-ec.org
 
\bibitem{DESI}
 http://desi.lbl.gov/cdr/
 
 \bibitem{eBOSS}
 http://www.sdss.org/surveys/eboss/
  
\bibitem{Creminelli:2008wc} 
  P.~Creminelli, G.~D'Amico, J.~Norena and F.~Vernizzi,
  ``The Effective Theory of Quintessence: the w<-1 Side Unveiled,''
  JCAP {\bf 0902}, 018 (2009)
  [arXiv:0811.0827 [astro-ph]].

  \bibitem{EFTOr} 
  G.~Gubitosi,  F.~Piazza and F.~Vernizzi,
  ``The Effective Field Theory of Dark Energy,''
  JCAP {\bf 1302}, 032 (2013)
  [arXiv:1210.0201 [hep-th]].
   
\bibitem{Bloomfield:2012ff} 
  J.~K.~Bloomfield, E.~Flanagan, M.~Park and S.~Watson,
  ``Dark energy or modified gravity?  An effective field theory approach,''
  JCAP {\bf 1308}, 010 (2013)
  [arXiv:1211.7054 [astro-ph.CO]].
  
  \bibitem{GLPV} 
 J.~Gleyzes, D.~Langlois, F.~Piazza and F.~Vernizzi,
 ``Essential Building Blocks of Dark Energy,''
 JCAP {\bf 1308}, 025 (2013)
 [arXiv:1304.4840 [hep-th]].
  
\bibitem{Bloomfield:2013efa} 
  J.~Bloomfield,
  ``A Simplified Approach to General Scalar-Tensor Theories,''
  JCAP {\bf 1312}, 044 (2013)
  [arXiv:1304.6712 [astro-ph.CO]].
  
\bibitem{PV} 
  F.~Piazza and F.~Vernizzi,
  ``Effective Field Theory of Cosmological Perturbations,''
  Class.\ Quant.\ Grav.\  {\bf 30}, 214007 (2013)
  [arXiv:1307.4350].
  
  \bibitem{Tsujikawa:2014mba} 
  S.~Tsujikawa,
  ``The effective field theory of inflation/dark energy and the Horndeski theory,''
  Lect.\ Notes Phys.\  {\bf 892}, 97 (2015)
  [arXiv:1404.2684 [gr-qc]].
  
  \bibitem{Gleyzes:2015rua} 
  J.~Gleyzes, D.~Langlois, M.~Mancarella and F.~Vernizzi,
  ``Effective Theory of Dark Energy at Redshift Survey Scales,''
  arXiv:1509.02191 [astro-ph.CO].
  
  \bibitem{eftcamb1} 
  N.~Frusciante, M.~Raveri and A.~Silvestri,
  ``Effective Field Theory of Dark Energy: a Dynamical Analysis,''
  JCAP {\bf 1402}, 026 (2014)
  [arXiv:1310.6026 [astro-ph.CO]].
  
  \bibitem{eftcamb2} 
  B.~Hu, M.~Raveri, N.~Frusciante and A.~Silvestri,
  ``Effective Field Theory of Cosmic Acceleration: an implementation in CAMB,''
  Phys.\ Rev.\ D {\bf 89}, no. 10, 103530 (2014)
  [arXiv:1312.5742 [astro-ph.CO]].
  
  \bibitem{eftcamb3} 
  M.~Raveri, B.~Hu, N.~Frusciante and A.~Silvestri,
  ``Effective Field Theory of Cosmic Acceleration: constraining dark energy with CMB data,''
  Phys.\ Rev.\ D {\bf 90}, no. 4, 043513 (2014)
  [arXiv:1405.1022 [astro-ph.CO]].
  
    \bibitem{Gergely:2014rna} 
  L.~\'A.~Gergely and S.~Tsujikawa,
  ``Effective field theory of modified gravity with two scalar fields: dark energy and dark matter,''
  Phys.\ Rev.\ D {\bf 89}, 064059 (2014)
  [arXiv:1402.0553 [hep-th]].
  
 \bibitem{Gleyzes:2015pma} 
  J.~Gleyzes, D.~Langlois, M.~Mancarella and F.~Vernizzi,
 ``Effective Theory of Interacting Dark Energy,''
  arXiv:1504.05481 [astro-ph.CO].


    \bibitem{horndeski}
G.~W.~Horndeski, 
Int.\ J.\ Theor.\ Phys. {\bf 10}, 363 (1974).

\bibitem{Deffayet:2009mn} 
  C.~Deffayet, S.~Deser and G.~Esposito-Farese,
  ``Generalized galileons: All scalar models whose curved background extensions maintain second-order field equations and stress-tensors,''
  Phys.\ Rev.\ D {\bf 80}, 064015 (2009)
  [arXiv:0906.1967 [gr-qc]].

\bibitem{Deffayet:2011gz} 
  C.~Deffayet, X.~Gao, D.~A.~Steer and G.~Zahariade,
  ``From k-essence to generalised Galileons,''
  Phys.\ Rev.\ D {\bf 84}, 064039 (2011)
   [arXiv:1103.3260 [hep-th]].
             
       
      \bibitem{NRT}
A.~Nicolis, R.~Rattazzi and E.~Trincherini,
  ``The galileon as a local modification of gravity,''
  Phys.\ Rev.\ D {\bf 79}, 064036 (2009)
  [arXiv:0811.2197 [hep-th]].
     
     \bibitem{Bellini:2015xja} 
  E.~Bellini, A.~J.~Cuesta, R.~Jimenez and L.~Verde,
  ``Constraints on deviations from ${\Lambda}$CDM within Horndeski gravity,''
  arXiv:1509.07816 [astro-ph.CO].
         
  
 \bibitem{pheno} 
  F.~Piazza, H.~Steigerwald and C.~Marinoni,
  ``Phenomenology of dark energy: exploring the space of theories with future redshift surveys,''
  JCAP {\bf 1405}, 043 (2014)
  [arXiv:1312.6111 [astro-ph.CO]].
  
  \bibitem{pheno2} 
  L.~Perenon, F.~Piazza, C.~Marinoni and L.~Hui,
  ``Phenomenology of dark energy: general features of large-scale perturbations,''
  JCAP {\bf 1511}, no. 11, 029 (2015)
   [arXiv:1506.03047 [astro-ph.CO]].
  
    \bibitem{BS} 
  E.~Bellini and I.~Sawicki,
  ``Maximal freedom at minimum cost: linear large-scale structure in general modifications of gravity,''
  JCAP {\bf 1407}, 050 (2014)
   [arXiv:1404.3713 [astro-ph.CO]].


\bibitem{Gleyzes:2014rba} 
  J.~Gleyzes, D.~Langlois and F.~Vernizzi,
  ``A unifying description of dark energy,''
  Int.\ J.\ Mod.\ Phys.\ D {\bf 23}, no. 13, 1443010 (2015)
  [arXiv:1411.3712 [hep-th]].
  
  \bibitem{Gleyzes:2014qga} 
  J.~Gleyzes, D.~Langlois, F.~Piazza and F.~Vernizzi,
  ``Exploring gravitational theories beyond Horndeski,''
  JCAP {\bf 1502}, 018 (2015)
   [arXiv:1408.1952 [astro-ph.CO]].

  
  \bibitem{Aubourg:2014yra} 
  \'E.~Aubourg, S.~Bailey, J.~E.~Bautista, F.~Beutler, V.~Bhardwaj, D.~Bizyaev, M.~Blanton and M.~Blomqvist {\it et al.},
  ``Cosmological implications of baryon acoustic oscillation (BAO) measurements,''
  arXiv:1411.1074 [astro-ph.CO].

  
  \bibitem{Cline:2003gs} 
  J.~M.~Cline, S.~Jeon and G.~D.~Moore,
  ``The Phantom menaced: Constraints on low-energy effective ghosts,''
  Phys.\ Rev.\ D {\bf 70}, 043543 (2004)
  [hep-ph/0311312].

  \bibitem{Kunz:2007rk} 
  M.~Kunz,
  ``The dark degeneracy: On the number and nature of dark components,''
  Phys.\ Rev.\ D {\bf 80}, 123001 (2009)
  [astro-ph/0702615].

  \bibitem{darkdeg} 
  M.~Kunz, S.~Nesseris and I.~Sawicki,
  ``Using dark energy to suppress power at small scales,''
  Phys.\ Rev.\ D {\bf 92}, no. 6, 063006 (2015)
  [arXiv:1507.01486 [astro-ph.CO]].
  
\bibitem{BPV} 
  J.~Beltran Jimenez, F.~Piazza and H.~Velten,
  ``Piercing the Vainshtein screen with anomalous gravitational wave speed: Constraints on modified gravity from binary pulsars,''
  arXiv:1507.05047 [gr-qc].
  

\bibitem{Adams:2006sv} 
  A.~Adams, N.~Arkani-Hamed, S.~Dubovsky, A.~Nicolis and R.~Rattazzi,
  ``Causality, analyticity and an IR obstruction to UV completion,''
  JHEP {\bf 0610}, 014 (2006)
  [hep-th/0602178].
  
\bibitem{Hojjati:2011ix}
  A.~Hojjati, L.~Pogosian and G.~B.~Zhao,
  ``Testing gravity with CAMB and CosmoMC,''
  JCAP {\bf 1108} (2011) 005
  [arXiv:1106.4543 [astro-ph.CO]].

\bibitem{VIPERS}
  J.~Bel, C.~Marinoni, B.~R.~Granett, L.~Guzzo, J.~A.~Peacock, and E.~Branchini {\it et al.},
  ``The VIMOS Public Extragalactic Redshift Survey (VIPERS) : $\Omega_{\rm m_0}$ from the galaxy clustering ratio measured at $z \sim 1$,''
  Astron.\ Astrophys.\  {\bf 563}, A37 (2014)
  [arXiv:1310.3380 [astro-ph.CO]].
  
  \bibitem{MGCAMB2011}
  A.~Hojjati, L.~Pogosian, G.~Zhao,
  'Testing gravity with CAMB and CosmoMC',
  JCAP {\bf 005}, 1108 (2011)
  
\bibitem{Adam:2015rua}
  R.~Adam {\it et al.} [Planck Collaboration],
  ``Planck 2015 results. I. Overview of products and scientific results,''
  arXiv:1502.01582 [astro-ph.CO].
  
\bibitem{Aghanim:2015xee}
  N.~Aghanim {\it et al.} [Planck Collaboration],
  ``Planck 2015 results. XI. CMB power spectra, likelihoods, and robustness of parameters,''
  [arXiv:1507.02704 [astro-ph.CO]].
  
\bibitem{bennett2012}
  C.~L.~Bennett {\it et al.} [WMAP Collaboration],
  ``Nine-Year Wilkinson Microwave Anisotropy Probe (WMAP) Observations: Final Maps and Results,''
  Astrophys.\ J.\ Suppl.\  {\bf 208} (2013) 20
  doi:10.1088/0067-0049/208/2/20
  [arXiv:1212.5225 [astro-ph.CO]].
  
  
  \bibitem[{{Haslam} {et~al.}(1982){Haslam}, {Stoffel}, {Salter}, \&
  {Wilson}}]{haslam1982}
{Haslam}, C., {Stoffel}, H., {Salter}, C.~J., \& {Wilson}, W.~E., {A 408\,MHz
  all-sky continuum survey. II - The atlas of contour maps }. 1982, Astronomy
  and Astrophysics Supplement Series, 47, 1
  
\bibitem{Lewis:2002ah} 
  A.~Lewis and S.~Bridle,
  ``Cosmological parameters from CMB and other data: A Monte Carlo approach,''
  Phys.\ Rev.\ D {\bf 66}, 103511 (2002)
  [astro-ph/0205436].
\bibitem{Lewis:2013hha} 
  A.~Lewis,
  ``Efficient sampling of fast and slow cosmological parameters,''
  Phys.\ Rev.\ D {\bf 87} (2013) 10,  103529
  doi:10.1103/PhysRevD.87.103529
  [arXiv:1304.4473 [astro-ph.CO]].

\bibitem{Gleyzes:2014dya} 
  J.~Gleyzes, D.~Langlois, F.~Piazza and F.~Vernizzi,
  ``Healthy theories beyond Horndeski,''
  Phys.\ Rev.\ Lett.\  {\bf 114}, no. 21, 211101 (2015)
    [arXiv:1404.6495 [hep-th]].
  
\bibitem{Simpson:2012ra}
  F.~Simpson {\it et al.},
  Mon.\ Not.\ Roy.\ Astron.\ Soc.\  {\bf 429} (2013) 2249
  doi:10.1093/mnras/sts493
  [arXiv:1212.3339 [astro-ph.CO]].
  
\bibitem{Pogosian:2016pwr}
  L.~Pogosian and A.~Silvestri,
  arXiv:1606.05339 [astro-ph.CO].
  
\end{thebibliography}
\end{document}